\documentclass[onecolumn]{IEEEtran}
\iftrue
\usepackage{graphicx}
\usepackage{amsmath}
\usepackage{subfigure}
\usepackage{algorithmic,algorithm}

\fi

 \title{Resilient Networking in Formation Flying UAVs }

 \author{  Lebsework Negash\footnote{Assistant Professor, Addis Ababa Institute of Technology, Addis Ababa University, lebsework@aait.edu.et} ~and ~Han-Lim Choi\footnote{Associate Professor, KAIST, hanlimc@kaist.ac.kr}}

\newtheorem{theorem}{Theorem}

 \newtheorem{definition}{Definition}
 \newtheorem{lemma}{Definition}
  \newtheorem{prop}{Proposition}

\begin{document}

\maketitle

\begin{abstract}
The threats on cyber-physical system have changed much into a level of sophistication that elude the traditional security and protection methods. This work addresses a proactive approaches to the cyber security of a formation flying UAVs. A resilient formation control of UAVs in the presence of non-cooperative (defective or malicious) UAVs is presented. Based on local information a resilient consensus in the presence of misbehaving nodes is dealt with fault-tolerant consensus algorithm. In the proposed framework, a graph-theoretic property of network robustness conveying the notion of a direct information exchange between two sets of UAVs in the network is introduced to analyze the behavior and convergence of the distributed consensus algorithm. A distributed control policy is developed to maintain the network connectivity threshold to satisfy the topological requirement put forward for the resiliency of the  consensus algorithm. Numerical examples are presented to show the applicability of the proactive approach used in dealing with the cyber attack treat on a formation flying UAVs.
\end{abstract}

\section {Introduction}

Recent advances in UAV capabilities for operating in an autonomous mode using high-end computing and communication infrastructure on board have spurred wide interest from law enforcement to commercial sectors in deploying a large number of UAVs. These make UAVs ideal candidates for coordinated tasks when it is not possible to perform the task single-handedly and efficiently \cite{miller2006mini,goodrich2008supporting,doherty2007uav}. Such tasks include search and rescue missions, or law enforcement activities such as border patrol and drug traffic monitoring~ \cite{waharte2010supporting,rudol2008human}. These tasks rely on the cooperative control of multiple-UAVs and their interaction with the environment with all its uncertainties.
Unlike a more centrally controlled system, distributed systems inherently have many vulnerable points of component failures, entry points for malicious attacks or intrusion. Hence, it is of increasing importance to guarantee trustworthy computation in the presence of misbehaving agents. A broken or a compromised node that fails to faithfully compute its values or a compromised sensor or communication network among the agents are some of the examples. To deal with this, the in-network computation algorithm needs to withstand a subset of failed or compromised nodes with some level of acceptable degraded performance. Such networked system referred as a resilient network to adversaries or faulty nodes.

In distributed networks, one of the important objectives is achieving consensus on estimated/computed variables of interest in a distributed manner~\cite{ren2005consensus, ren2007infoconsensus, olfati2007consensus, fraser2012hyper}. Here, individual nodes interact locally to accomplish some global objective. But as the interconnection of the networks grows in size and complexity, they become increasingly vulnerable to node or communication failure. The issue of agreement among nodes in the presence of adversaries has been extensively studied in the area of distributed computing \cite{lynch1996distributed, pease1980reaching}. In the context of network control systems, resilient protocols for consensus in the presence of malicious, Byzantine, non-colluding and other threat models have been studied recently, and it is observed that the resilience of various consensus algorithms is tied strongly to the underlying topology of networks.

In this work, we studied resilient consensus problem for a network of UAVs with a second-order dynamics where the maximum number of misbehaving UAVs in the network is known. A W-MSR algorithm is used for well-behaving UAVs to fly in a formation but the convergence of a resilient consensus guaranteed on the  $(2F+1)$-robustness of the network topology \cite{leblanc2013resilient}. Since computing the exact robustness of a network is a coNP-complete problem, a relationship with the algebraic connectivity of the graph as the r-robustness lower limit is established. A communication management controller designed so that the controller gives a local direction to move in order to improve or maintain the connectivity of the network measured by the algebraic connectivity ($\lambda_2$). Our work differs from \cite{saulnier2017resilient, stump2008connectivity} in that computing algebraic connectivity of graph and its corresponding vector need a global knowledge of the network which is not available to the individual UAV node. Hence, for our framework of distributed communication management controller both  $\lambda_2$ and its corresponding eigenvector, $ v_2$,  are estimated via power iteration method at each node. Numerical examples are presented showing the effectiveness and resilience of the consensus algorithm maintaining the UAVs in their route despite different model of misbehaving/compromised UAVs in the network.

The rest of the paper is organized as follows. A graph theory based formation control algorithm is presented in section II. Section III describes a formal definition of resilient network and topological criteria for WMSR algorithm to reach consensus. Section IV describes graph robustness control through algebraic connectivity of the graph. Simulation results of resilient network in formation flying UAVS presented in section V. Section VI concludes the paper.

\section{Formation Control of UAVs} \label{sec:formation}
In this section, the formation control from \cite{fax2004information} is adapted so that it will suit the specific purpose of this work and the dynamics of the UAVs considered. Starting from here, in a formation flying the UAVs will be referred to agents or nodes interchangeably.

Consider $N$ UAVs coordinating themselves to achieve a pre-specified formation defined by relative positions with respect to each other. To describe the interaction architecture in a formal manner, consider an undirected graph $\mathcal{G}=(\mathcal{V},\mathcal{E})$, where the set of nodes $\mathcal{V}=\{{v_1,...v_N}\}$ and the set of edge, $\mathcal{E}\subseteq\mathcal{V}\times\mathcal{V}$. The neighbors of UAV $i$ are denoted by $\mathcal{N}_i=\{j \in\mathcal{V}:(i,j)\in\mathcal{E}\}.$ Every node represents a UAV and the edges correspond to inter-vehicle communication.
The adjacent matrix $\mathcal{A} \in \{0,1\}^{N \times N}$ represents the adjacency relationship in graph $\mathcal{G}$ with an element $a_{ij}=1$ if $(v_i,v_j )\in\mathcal{E}$ and $a_{ij}=0$ otherwise. The neighbor of agent $i$, denoted $\mathcal{N}_i$ is the set of agents such that $a_{ji} = 1$. The graph Laplacian is defined as
\begin{equation}
	L_{\mathcal{G}} = \mathcal{D} - \mathcal{A}\label{eq:L_g}
\end{equation}
where $\mathcal{D}$ is a diagonal matrix with $d_{ii} $ representing the cardinality of $ \mathcal{N}_i$.

The motion of each UAV in $d$-dimensional Euclidean space is modeled as a second order system:
\begin{equation}
	\dot{x}_i =A_i x_i+B_i u_i, \qquad x_i \in\mathbf{R}^{n} \label{eq:x_i}
\end{equation}
where the state variable $x_i$ consists of the configuration variables (i.e., position) and their derivatives (i.e., velocity) and the control input $u_i$ represents the acceleration commands.

\begin{equation}
	u_i = -k_3 v_i+  \frac{1}{|\mathcal{N}_i | }\sum_{j \in \mathcal{N}_i} a_{ij}\left[ (x_i - h_i ) - (x_j - h_j) \right ], \qquad \forall i \in \{1, \dots, N \} \label{eq:u_i}
\end{equation}

where $h_i =  \tilde{h}_i \otimes [1~0]^T$, with some feedback gain $K_i \in \mathbf{R}^{n \times 2n} $. The cardinality $|\mathcal{N}_i |$ is used for normalization \cite{fax2004information,lafferriere2005decentralized}. Because UAV motion is modeled as a second-order system with acceleration input, $K_i$ takes the form of $K_i = I_n \otimes \left[ k_{i, pos}, k_{i, vel} \right]$.
With the state equation in (\ref{eq:x_i}) and control input in (\ref{eq:u_i}), the overall closed-loop dynamics of the fleet can be written as:
\begin{equation}
	\dot{x} = A x + B K L ( x- h) \label{OVR_allFormation}
\end{equation}
with the overall state $x = [x_1^T, \dots, x_N^T]^T$ and desired formation $h = [h_1^T, \dots, h_N^T]^T$, where
$A = I_N \otimes A_i, ~B = I_N \otimes B_i,~K = I_N \otimes K_i, ~L = L_{\mathcal{G}} \otimes I_n.
$

\section{Resilient Network Topology and Formation Control }

A fundamental challenge of in-network computation and reaching consensus in the quantities of interest is that it must be calculated using only local information. Reaching consensus in the presence of misbehaving or faulty nodes resiliently has been  shown that the correctly behaving nodes can reach consensus overcoming up to $F$ misbehaving nodes if the network connectivity is at least $2F+1$ \cite{lynch1996distributed,sundaram2011distributed,leblanc2013resilient,pasqualetti2012consensus}. But these algorithms either require those normal nodes to have at least nonlocal information or assume that the network is complete.

Consider a time varying network modeled by a digraph $\mathcal{G}(t)=(\mathcal{V},\mathcal{E}(t))$, where the set of nodes $\mathcal{V}=\{{v_1,...v_N}\}$ and the set of edge, $\mathcal{E}(t)\subseteq\mathcal{V}\times\mathcal{V}$. The neighbors of node $i$ are denoted by $\mathcal{N}_i=\{j \in\mathcal{V}:(i,j)\in\mathcal{E}(t)\}$. Base on the nomenclature in \cite{leblanc2013resilient, zhang2012robustness}, let us assume that the nodes are partitioned into a set of $\mathcal{N}_n$ normal nodes $\mathcal{N}_n= \{1,2,...N_n\}$ and $\mathcal{N}_a$ a set of adversary nodes  $\mathcal{N}_a=\{N_n +1,N_n+2,...N\}$. The nodes are assumed to have unique identifiers that forms a totally ordered set.
\subsection{Connectivity } \label{sec:connectivity}
The UAVs equipped with wireless communication system that allows point-to-point communication between individual UAVs. The time varying edge $\mathcal{E}(t)$ represents the connection between pair of $(i,j)$ agents. The link modeled as range-dependent \cite{de2006decentralized,stump2008connectivity}, with a quality varying between $0$ and $1$ based on the the communication disk model. The strength of the connection among agents $(i,j)$, which decays exponentially with the distance, will be:
\begin{equation} \label{eq: communication function}
	f_{ij}(x)=
	\begin{cases}
		1, & \quad  \lVert x_i-x_j \lVert < \rho\\
		0, & \quad  \lVert x_i-x_j \lVert \geq R\\
		exp(\frac{-\gamma(\lVert x_i-x_j \lVert-\rho)}{R-\rho}) & \quad\text{otherwise}
	\end{cases}
\end{equation}
\noindent
where $R$ defines the maximum communication range radius and $\rho$ is distance where the communication signal strength is optimal.
\begin{figure}[h]
	\centering
	\includegraphics[width=0.35\columnwidth]{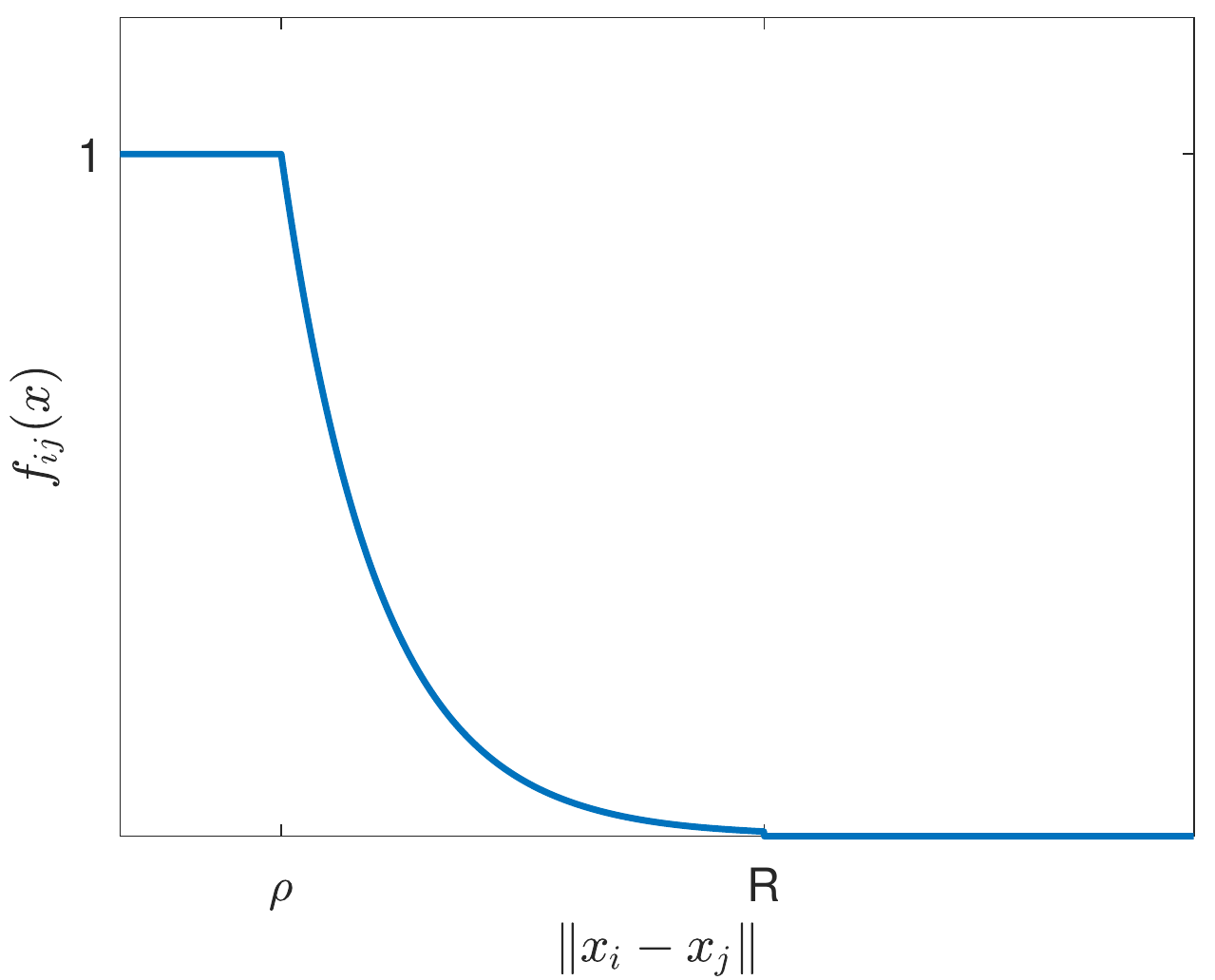}
	\caption{Communication strength between UAVs $i$ and $j$ as a function of distance between them. }
	\label{fig1}
\end{figure}

The adjacent matrix $\mathcal{A}(x)$ represents the adjacency relationship strength in graph  $\mathcal{G}$ with an element $a_{ij}=f_{ij}$ representing the strength of the connection given by the edge  $(v_i,v_j )\in\mathcal{E}(x)$. The weighted Laplacian, $L_{\mathcal{G}}(x)$, is given as :
\begin{equation}
	L_{\mathcal{G}}(x) = \mathcal{D}(x) - \mathcal{A}(x)\label{eq:L_g(x)}
\end{equation}

where $\mathcal{D}(x)$ is a diagonal matrix consisting of the row-sum of $\mathcal{A}(x)$. Since $L_{\mathcal{G}}$ is positive-semi-definite and symmetric, with all its eigenvalues are real, and its special structure ensures that its smallest eigenvalue is zero with the corresponding eigenvector of all ones, $\textbf{1}$. The second smallest eigenvalue of the Laplacian known as the algebraic connectivity (or Fiedler value), denoted as $\lambda_2$, indicates the overall connectedness of a graph. Hence, the magnitude of the algebraic connectivity reflects how well connected the overall graph is.

\subsection{System Dynamics}
Suppose each UAV node $i \in \mathcal{V} $  starts with initial value $x_i[0] \in \mathbf{R}^n $ where $x(t)=[x_{\mathcal{N}_n}^T,x_{\mathcal{N}_a}^T]$ and each normally behaving nodes interact with their neighbors and updates their values according to some nominal rule which is modeled as
\begin{equation}
	x_i(t+1)=f_i(x_{\mathcal{N}_n} ,x_{\mathcal{N}_a},t), i\in \mathcal{N}_i
\end{equation}
where each function $f_i(.)$ is a predefined function so that the normal nodes reach consensus. The misbehaving nodes would try to influence the well-behaving nodes, thus $f_i(.)$ should be designed in a way the well-behaving nodes should eliminate or reduced their effect with out prior knowledge of their identities. Here, the scope of threat model defines the topological assumptions placed on the adversaries. Definitions adapted from \cite{leblanc2013resilient}.
\begin{definition}
	($F$-total set): A set $\mathcal{S} \subset \mathcal{V}$ is $F$-total if it contains at most $F$ nodes in the network, i.e.,  $ \mid \mathcal{S} \mid \leq F,\, F \in  \mathbf{Z}_{\geq 0}$ .
\end{definition}
\begin{definition}
	($F$-local set): A set $\mathcal{S} \subset \mathcal{V}$ is $F$-local if it contains at most $F$ nodes in the neighborhood of the other nodes, i.e.,  $\mid \mathcal{V}_i  \bigcap  \mathcal{S} \mid \leq F,\,  \quad \forall i \in \mathcal{V} \backslash \mathcal{S},\, F \in  \mathbf{Z}_{\geq 0}$ .
\end{definition}
Note: $\mathbf{Z}_{\geq 0}$ is a set of integer greater than or equal to zero.

\noindent
A set of adversary nodes is $F$-totally bounded or $F$-locally bounded if it is an $F$-total set or $F$-local set respectively.

\begin{definition}
	($r$-reachable subset): The subset $\mathcal{S} \subset \mathcal{V}$ is said to be $r$-reachable if $\exists i \in \mathcal{S}$ such that $\mid \mathcal{V} \backslash \mathcal{S} \mid  \geq r$, where $r \in \mathbf{Z}_{\geq 0} $, i.e., if it contains a node that has at least $r$ neighbors outside that set.
\end{definition}
\noindent
A set $\mathcal{S}$ is $r$-reachable if it contains a node that has at least $r$ neighbors outside of $\mathcal{S}$. The parameter $r$ quantifies the redundancy of information flow from nodes outside of $\mathcal{S}$ to some node inside $\mathcal{S}$. Intuitively, the $r$-reachability property captures the idea that some node inside the set is influenced by a sufficiently large number of nodes from outside the set.
\begin{definition}
	($r$-robust graph): A graph $\mathcal{G}$ is said to be $r$-robust if for every pair of nonempty disjoint subset of $\mathcal{V}$, at least one of the subset is $r$-reachable.
\end{definition}
\begin{figure}[h]
	\centering
	\includegraphics[width=0.4\columnwidth]{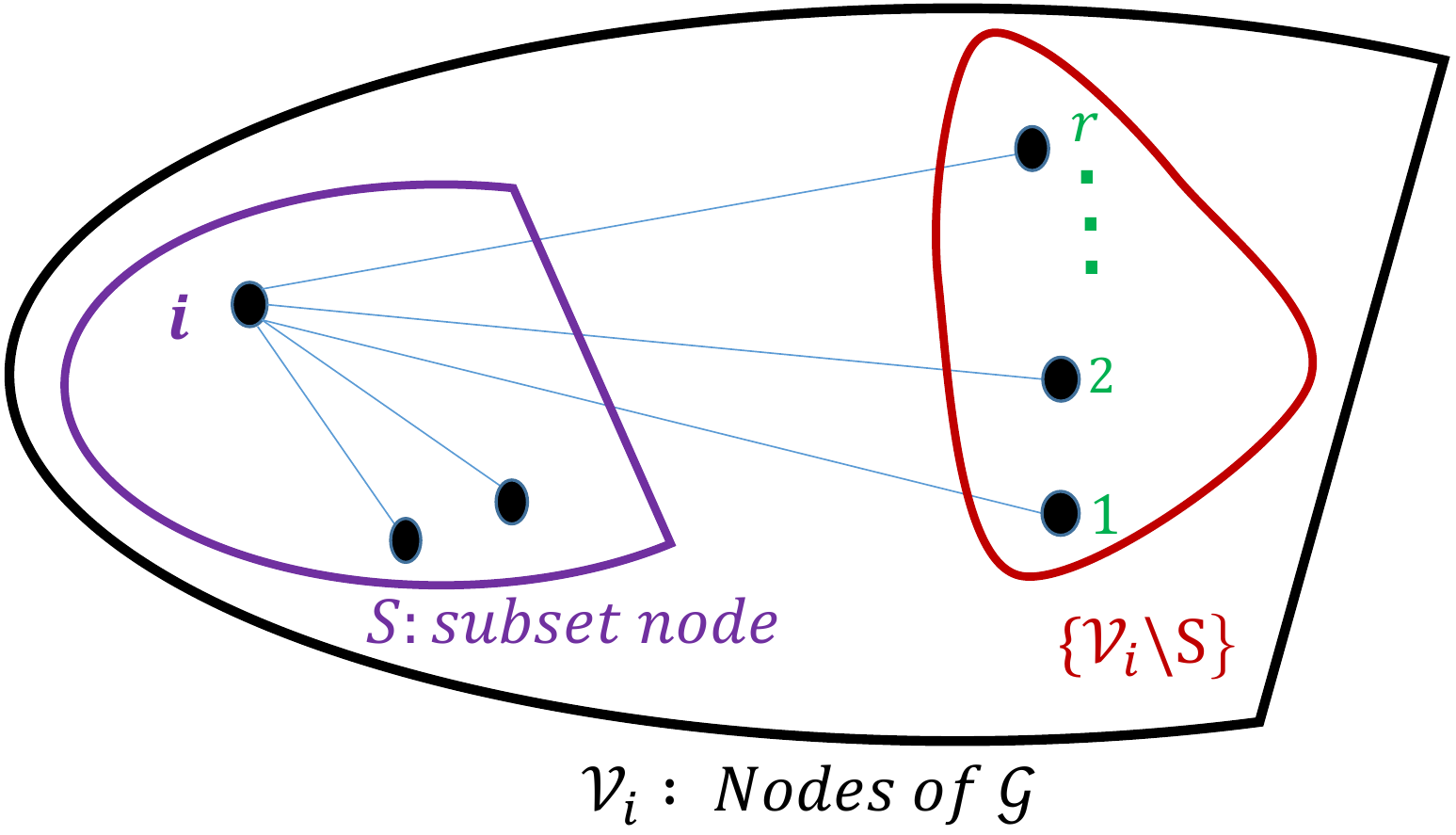}
	\caption{A subset of nodes $S$ is accessible from outside by $r$- number of nodes from  $ \{ \mathcal{V}_i    \setminus S \}$.
	}
	\label{r-rechable}
\end{figure}

In summary, an $r$-reachable set contains a node that has $r$ neighbors outside that set, and a $r$-robust graph has the property that no matter how one chooses two disjoint nonempty sets, at least one of those sets is $r$-reachable. Therefore, if a graph $\mathcal{G}$ is $r$-robust, then it is at least $r$-connected and has minimum degree at least $r$ \cite{leblanc2013resilient}.

\section{Resilient Asymptotic Consensus} \label{resilient Asymptotic cons defn}
\subsection{W-MSR Algorithm}
The Weighted-Mean-Subsequence-Reduced (W-MSR) algorithm is a modified version of the linear consensus algorithms. The linear consensus protocol operates by updating each normal agent as $x(t)$ in (\ref{eq:x_i})  with control input which uses the relative position and velocity with its neighbors.

\begin{equation} \label{eq:control enput W-MSR}
	\begin{array}{ll}
		u_i(t) = -\kappa_i v_i(t) + \sum\limits_{j\in N_i} a_{ij}(t)\left[ (x_j(t)-x_i(t))	+ \gamma (v_j(t)-v_i(t)) \right]
	\end{array}
\end{equation}
where $a_{ij}(t)$ is the $(i,j)$ entry of the adjacency matrix of the corresponding time varying Laplacian of the communication graph $\mathcal{G}(x(t))$ as the states are a function of time.

The W-MSR algorithm with parameter $F\geq 0$ alters the above protocol by having each agent remove state values that are relatively extreme compared to the rest of the agent’s in-neighbor set and its own state. Specifically, it works as follows \cite{leblanc2013resilient,dibaji2017resilient}

\begin{algorithm}[h]
	\caption{ Weighted-Mean-Subsequence-Reduced (W-MSR) algorithm }
	\label{alg:W-MSR}
	\begin{enumerate}
		\item At each time-step $t$, each normal node $ i \in\mathcal{N} $ obtains the {}{position} values of its neighbors, and forms a sorted list.
		\item If there are less than $F$ {}{position} values larger than its own {}{position}, then normal node $i$ removes all {}{position} values that are larger than its own. Otherwise, it removes the largest $F$ {}{position} values in the sorted list. Likewise, if there are less than $F$ {}{position} values smaller than its own {}{position} value, then node $i$ removes all {}{position} values that are smaller than its own. Otherwise, it removes the smallest $F$ {}{position}.
		\item let $\mathcal{P}_i(t)$ is the set of nodes whose {}{position} were removed by normal node $i$ in step 2 at time step $t$. Each normal node $i$ applies the state update  in (\ref{eq:x_i}) applying the control rule in (\ref{eq:control enput W-MSR}) (of each dimension), except $\mathcal{N}_i(t)$ replaced by $\mathcal{N}_i(t) \backslash \mathcal{P}_i(t) $.
		
		\begin{equation} \label{eq:Linear_cons_W-MSR}
		 u_i(t) = -\kappa_i v_i(t) + \sum\limits_{j \in \mathcal{N}_i(t) \backslash \mathcal{P}_i(t)} a_{ij}(t)\left[ (x_j(t)-x_i(t))	+ \gamma (v_j(t)-v_i(t)) \right] 			
		\end{equation}
		
		(i.e., \textit{apply control input (\ref{eq:control enput W-MSR}) by substituting $a_{ij}(t)=0$ for the incoming edges $(i,j)\in\mathcal{E}(t) $ of agents ignored in step 2.})
	\end{enumerate}
\end{algorithm}

\begin{theorem} \label{theorem:w-msr (2F+1) robustness}
	\cite{leblanc2013resilient} Consider a time varying digraph $\mathcal{G}(x(t))=(\mathcal{V},\mathcal{E}(x(t))$, where each node with second-order dynamics updates its value according to W-MSR algorithm \ref{alg:W-MSR} with a parameter $F$. Under $F$-local/total malicious model, resilient asymptotic consensus is achieved if the topology of the network is $(2F+1)$ robust.
	
	\noindent
	For proof, the reader kindly advised to look into there references \cite{leblanc2013resilient,dibaji2017resilient}.
\end{theorem}

\subsection{Maintaining Robust and Resilient Network}
The topological property, $(2F+1)$ robustness of a network, is the sufficient condition for the W-MSR algorithm to guarantee consensus. However, determining whether a given graph is $r$-robust for any $r \geq 2$ is coNP-complete \cite{zhang2012robustness,zhang2015notion}. In case of a dynamic network, such as a team of UAVs' network, to maintain the resiliency of the network, the controller should keep the robustness of the dynamic network in check every time step. This will become intractable. It is better to resort to a more intuitive and computable parameter of the network relatable to r-robustness.

\subsection{Lower bound on r-robustness}
The work in \cite{shahrivar2015robustness} proved that, the graph Cheeger constant \footnote{Cheeger constant or isoperimetric number of a graph is a numerical measure of whether or not a graph has a ``bottleneck".} (also called isoperimetric number $i(\mathcal{G})$) is a lower bound to $r$.
\begin{lemma} \label{cheeger-robust}
	\cite{shahrivar2015robustness} For a given graph $\mathcal{G}$ and its isopermimetric number, $i(\mathcal{G})$, if  $i(\mathcal{G}) > r-1$, then the graph is at least r-robust.
\end{lemma}
However, in \cite{kaibel2004expansion} while working on expansion of graphs, and in \cite{arias2012normalized} the authors proved that computing Cheeger constant of a graph is NP-hard and motivate the use of spectral methods. But earlier the authors in  \cite{chung1997spectral,fiedler1973algebraic} showed that the Cheeger constant is lower bounded by the algebraic connectivity of the graph.

\begin{lemma}\label{cheeger-lambda_2}
	\cite{chung1997spectral} $\frac{i(\mathcal{G})^2}{2d_{max}} \leq \lambda_2(L_{\mathcal{G}}) \leq 2i(\mathcal{G}) $, where the graph Laplacian real eigenvalues can be ordered sequencially $0=\lambda_1(L_{\mathcal{G}}) \leq \lambda_2(L_{\mathcal{G}})\leq ...\leq \lambda_n(L_{\mathcal{G}})\leq 2d_{max} $, $d_{max}$ is the maximum degree of a node in graph $\mathcal{G}$.
\end{lemma}

\noindent
Using Lemmas \ref{cheeger-robust} and \ref{cheeger-lambda_2}, lead to a lower bound on the algebraic connectivity to ensure r-robustness.

\begin{equation}
	2(r-1) < \lambda_2
\end{equation}

\noindent
Taking the W-MSR topological property for resilient convergence, ($2F+1$)- robustness, the convergence can be guaranteed as long as the graph connectivity is :
\begin{equation}\label{eq:connectivity bound}
	\lambda_2 > 4F
\end{equation}

\noindent
When this lower bound applied to regulate the connectivity of the network, it will satisfy the lower bound for the W-MSR algorithm to reach asymptotic consensus resiliently leading us to a well-dealt connectivity management in mobile robotics or communication management in wireless mobile sensors networks.

\section{Robustness Control through Graph Algebraic Connectivity} \label{sec:connectivity control}

The relationship between $\lambda_2$ and the overall connectivity of the graph can be used to formulate a control action either to preserve or increase the connectivity of the graph. The algebraic connectivity $(\lambda_2)$ is a function of the state of the entire graph where we can write $\lambda_2(L_{\mathcal{G}}(x))$, and it is a concave function of the graph Laplacian $(L_{\mathcal{G}})$ in the space $\textbf{1}^\perp$.
It is the infimum of a set of linear function in $L_{\mathcal{G}}$ \cite{sun2006fastest}:
\begin{equation}
	\lambda_2(L_{\mathcal{G}}(x))=\inf_{v \in \textbf{1}^\perp} \frac{v^TL_{\mathcal{G}}(x)v }{v^Tv}
\end{equation}
The function is not smooth, hence does not have gradient at each point. But in \cite{stump2008connectivity}, for distinct $\lambda_2$, i.e., $\lambda_2 \neq \lambda_3$, the derivative is shown that:
\begin{equation}
	\frac{\partial \lambda_2(x)}{\partial L_{\mathcal{G}}(x)} =\frac{v_2 v_2^T}{v_2^T v_2}
\end{equation}
where $v_2$ is eigenvector corresponding to the eigenvalue $\lambda_2$ and ${v_2 v_2^T}$ is a supergradient for $\lambda_2(L_{\mathcal{G}}(x))$ and can be used in maximizing the concave function. Though, $\lambda_2$ is concave with respect to $L_{\mathcal{G}}$, the Laplacian is nonlinear function of $x$. Using chain rule to linearize $\lambda_2$, derivative is taken with respect to each single element of $x$ :
\begin{equation}\label{eq:lambda_2 gradient}
	\frac{\partial \lambda_2(x)}{\partial x_{i,\alpha}}=\frac{\partial \lambda_2(x)}{\partial L_{\mathcal{G}}(x)}\frac{\partial L_{\mathcal{G}}(x)}{\partial x_{i,\alpha}} = Trace \left \{ \left [\frac{v_2 v_2^T}{v_2^T v_2}\right ]^T \left [\frac{\partial L(x)}{\partial x_{i,\alpha}} \right ]   \right \}
\end{equation}
The gradient, $ \frac{\partial \lambda_2(x)}{\partial x_{i,\alpha}}$, gives each agent local direction to move to improve or maintain connectivity of the graph. However, the decentralized control action of each agent with  information only about its neighbors, $\mathcal{N}_i$, is trying to increase the value of $\lambda_2$ which is a global knowledge of the graph property.

\section{Distributed Estimation of Algebraic Connectivity and its Eigenvector}
As it is shown in ( \ref{eq:lambda_2 gradient}), to generate a decentralized  control action to drive the agents to a more connected form, each individual agent need to compute $\lambda_2$ and $v_2$ in a decentralized way.
In our case, the global topology of the network is not available at each node (UAV) has access only to the information of its neighbors. Consequently, the second smallest eigenvalue and its corresponding eigenvector of the network cannot be directly computed. Hence, both variables related to the Laplacian matrix of the graph should be estimated or recovered distributively, i.e., by using only local information of each node in the network.

Various works have been proposed to estimate these network variables distributively. In \cite{franceschelli2009decentralized}, eigenvalues of Laplacian matrix are estimated using fast Fourier transform (FFT) by constructing distributed oscillators whose states oscillate at frequencies corresponding to the eigenvalues of graph Laplacian and then the agents use FFT on their state to identify eigenvalues, however, the FFT technique is not appropriate for real-time implementation and for handling switching topologies. The estimate of the second-smallest eigenvalue of the Laplacian matrix is used by De Gennaro and Jadbabaie \cite{de2006decentralized} to implement an optimization algorithm, that aims at increasing the value of the algebraic connectivity of the graph. Tran and Alain \cite{kibangou2015distributed} posed the problem as a constrained consensus problem formulated two-fold, as a direct and indirect approach yields non-convex optimization problem and convex optimization after adequate re-parametrization respectively. They solved both problems in a distributed way by means of a method of Lagrange multipliers and sub-gradient algorithm respectively. In our case, the main drawback of connectivity maintenance based on optimization algorithm arises from the difficulties in formally guaranteeing connectivity maintenance in the presence of conflicting objective that arises as the UAVs system are required to increase connectivity while performing formation flight \cite{sabattini2013decentralized}.

Exploiting a gradient-based algorithm appears as an attractive solution in order to provably guarantee connectivity maintenance. Here in our Laplacian spectral estimation relied on a Power Iteration method and its variation. Power Iteration starts with an initial vector, which may be an approximation to the dominant vector, then repeatedly multiplies it by a matrix and normalize it. The vector converges to the eigenvector associated with the dominant eigenvalue \cite{alston1975householder}. To get a particular eigenvalue, the original matrix deflated and the process repeated again. The main limitation of this approach is it requires normalization and orthonormalization of a vector at each step which requires global knowledge but can be solved by replacing average iteration as in Yang et.al \cite{yang2010decentralized}. But in both cases, the authors only concentrate on estimating the connectivity while we need also the corresponding eigenvector to use them in the above gradient-base control algorithm to control the connectivity of the UAVs.

\subsection{Decentralized Power Iteration}
Let $P$ be a deflated version of Perron Matrix \cite{olfati2007consensus} of the Laplacian $L_{\mathcal{G}}$
\begin{equation}\label{eq:Deflated matrix}
	\begin{split}
		P=D-\textbf{1 1}^T/n \quad \\ \text{where $D$ be not-deflated matrix}\quad  D=\textbf{I}-\alpha L_{\mathcal{G}}
	\end{split}
\end{equation}
The eignvalues of the Laplacian and the deflated version of the Perron matrix P are related by
\begin{equation}\label{eq:rela}
	\lambda_1(P)=0,\quad  \lambda_i(P)=1-\alpha \lambda_i(L_{\mathcal{G}}), \quad  \text{for } i\in\{2, ..., n\}
\end{equation}

The spectral radius $\rho(P)$ of $P$ is related to the algebraic connectivity $\lambda_2(L_{\mathcal{G}})$ by \cite{yang2010decentralized, aragues2012distributed, aragues2014distributed}

\begin{equation}\label{eq:deflimits}
	\lambda_2(L_{\mathcal{G}})=\frac{(1-\rho(P))}{\alpha},  \quad  \text{if}\quad  0 < \alpha < \frac{1}{\lambda_n(L_{\mathcal{G}})}
\end{equation}

Computing the spectral radius of matrix $P$ distributively will give the second smallest eigenvlue of the laplacian (\ref{eq:deflimits}). The spectral radius of a matrix defined as

\begin{equation}\label{eq:defSpecRadius1}
	\rho(P)=max\{ |\lambda_1(P)|,... ,|\lambda_n(P)| \}
\end{equation}
Gelfand's formula states that
\begin{equation}\label{eq:Gelfand's formula}
	\rho(P)=\lim_{k \to \infty} \lVert P^k \rVert^{1/k}_\infty
\end{equation}
Therefore, each UAV computes the $\infty$-norm, $\lVert . \rVert_\infty$, of the matrix $P^k$ which is the maximum absolute row sum of the matrix
\begin{equation}\label{eq:inf norm}
	\lVert P^k \rVert_\infty=\max\limits_{i \in v } \{ \lvert[P^k]_{i1}\rvert + ... + \lvert[P^k]_{in}\rvert  \}
\end{equation}
In this case then, each UAV can calculate the above matrix power using  max-consensus algorithm as each UAV $i$ knows the $i^{th}$ row of $P^k$. Hence, each UAV should compute the power $P^k$ of matrix $P$ which is not compatible with the graph. But, if the UAVs compute power of $D^k$ which is compatible\footnote{An $n\times n$ matrix $D$ is compatible  with $\mathcal{G}$,  $D_{ij}=0$ if $(i,j)\notin \mathcal{E}$ for $j\neq i$ and let $D_{ii}$ be equal or different than $0$.} with the graph, they can easily compute $P^k$ as shown below.  This is clear from the definition of $D$ below in (\ref{eq:defD})
\begin{equation}\label{eq:defD}
	D=\textbf{I}-\alpha L_{\mathcal{G}}=P+\textbf{11}^T/n
\end{equation}

The relationship between this two matrices are given in \cite{aragues2014distributed, yang2010decentralized}. Let's state it here the theorem.

\begin{theorem}
	\cite{aragues2014distributed} Consider a symmetric Laplacian matrix $L_{\mathcal{G}}$, let $D=\textbf{I}-\alpha L_{\mathcal{G}}$, with $\alpha >0$, and let $P=D-\textbf{11}^T/n$. The $k^{th}$ powers of matrices $P$ and $D$ are related as follows, for all $k\geq1$,
	\begin{equation}\label{eq:P_power_k}
		P^k=(D-\textbf{11}^T/n)^k=D^k-\textbf{11}^T/n
	\end{equation}
	For proof, the reader asked kindly to look into the reference.
	
\end{theorem}

The powers of $D^k$ of a matrix $D$ compatible with the graph can be computed in a distributed fashion by letting each UAV $i$ stores the $i^{th}$ row of $D^k$ and update each elements of the matrix, $[D^k]_{ij}$ for $j=1,...,n$ with its neighbor's data
\begin{equation}\label{eq:D_barUpndateRule}
	[D^{k+1}]_{ij}=\sum_{j'\in \mathcal{N}_i \cup \{i\}} D_{ij'}[D^k]_{j'j}
\end{equation}
since $D$ compatible with $\mathcal{G}$, $D_{ij'}=0$ for non-neighbor agents $j'\notin  \mathcal{N}_i \cup \{i\}$ and equation \ref{eq:D_barUpndateRule} becomes
\begin{equation}\label{eq:D_barUpndateRuleJen}
	[D^{k+1}]_{ij}=\sum_{j=1}D_{ij'}[D^k]_{j'j}
\end{equation}
Note that each UAV $i$ updates its variable using only its neighbors and its own values at each step $k$. The algorithm computes the exact $D^k$ (not an estimate). But observing (\ref{eq:D_barUpndateRule}), as each UAV $i$ receives $[D^k]_{j'j}$ from its neighbor $j'$, it needs to know to which agent $j$ this information belongs to while combining with its own $[D^k]_{ij}$. A more elaborative Algorithm is given in \cite{aragues2014distributed}, where we further use it to identify the number of UAVs in a network distributively.

Each agent will keep a list $Id_i(k)$ of the identifiers associated to each of its elements $[D^k]_{ij}$ and exchange these identifiers with their neighbors at each step $k$. As when  UAVs $i$ discovers in its neighbor's data $j'\in \mathcal{N}_i$ the identifier of a new UAV $j\in Id_{j'}(k)$, it initializes a new variable $[D^k]_{ij}$ and updates $Id_i(k)$ accordingly.

\begin{algorithm}
	\caption{Distributes Matrix Power computation} \label{euclid}
	\label{alg:D_barUpdate}
	Each node $i\in \mathcal{V}$ maintains a set of node identifiers $Id_i(k)$ and an estimate of $\hat{D}^k_{ij}$ of the $(i,j)$ entries of the $k^{th}$ of $D$, $[D^k]_{ij}$, associated to the nodes $j$ such that $j\in id_i(k)$
	\begin{enumerate}
		\item At $k=0$, each node $i\in \mathcal{V}$ initialize a single variable $[D^k]_{ij}$ and its identifier and send it to its neighbors $\mathcal{N}_i$
		\begin{equation}\label{eq:D_bar_intialization}
			D^k_{ij}=1 \quad id_i(0)=\{i\}
		\end{equation}
		\item  At each step $k \geq 1$, node $i$ first look for new nodes in the information $Id_{j'}(k)$ received from its neighbors $j' \in \mathcal{N}_i$ and updates its identifier $Id_i(k)$ .
		\begin{equation}\label{eq:Id_update}
			Id_i(k+1)=\bigcup_{j'\in \mathcal{N}_i \cup \{i\}} Id_{j'}(k)
		\end{equation}
		
		\item Then, node $i$ creates new variable $[\hat{D}^k]_{ij}$ initialized with $[\hat{D}^k]_{ij}=0$, for each recently discovered node $j$ in (2).
		\begin{equation}\label{eq:D_bar_neigh_update}
			j \in Id_i(k+1) \quad \text{and} \quad j \notin Id_i(k)\\
		\end{equation}
		
		\textbf{if} $Id_i(k+1)=Id_i(k)$   \textbf{then}\\
		\quad \text{number of UAV in the graph } $\mathcal{G}$ \text{ is} $n=\lvert Id_i(k)\rvert$;\\
		\quad Stop Algorithm 1 and go to Algorithm \ref{alg:conne_est};\\
		$\>$ \textbf{else}\\
		Continue with step 4\\

\item Finally, node $i$ updates all its variables $\hat{D}_{ij}(k)$,
\begin{equation}\label{eq:D_barUpndateRuleAlg}
	[\hat{D}^{(k+1)}]_{ij}=\sum_{j'\in \mathcal{N}_i \cup \{i\}, j\in Id_{j'}} [D]_{ij'} [\hat{D}^k]_{j'j}
\end{equation}
for $j \in Id_i(k+1)$, and sends to its neighbors these variables $[\hat{D}^{(k+1)}]_{ij}$  and the identifiers $Id_i(k+1) $.
\end{enumerate}
\end{algorithm}

The output of the algorithm $[\hat{D}^k]_{ij}$ at each UAV is an exact value (not an estimate) of the entries of $[D^K]_{ij}$ of $D^k$ associated to agents $j \in Id_i(k)$ which are at $k$ or less hops \cite{bullo2009distributed}. From (\ref{eq:P_power_k}), the aim is to calculate the power of $P^k=D^k-\textbf{11}^T/n $, hence the equivalent entries of $P^k$ is obtained by subtracting $1/n$ from the $[\hat{D}^k]_{ij}$. In addition the entries of $[\hat{D}^k]_{ij}$ for which $j \notin Id_i(k)$ is zero, therefore for each of these $n-\lvert Id_i(k) \rvert$ entries, the associated entries in $P^k$ equals $-1/n$.

\begin{prop} \label{Prp: max step to identify n}
Let each UAV $i$ run  Algorithm~\ref{alg:D_barUpdate} with $\mathcal{G}$ connected. The instant $k^*$ when UAV $i$ discovered the identifiers of all the UAVs in the network, then

\begin{equation}
Id_i(k)=Id_i(k-1) \quad \text{for} \quad \forall k>1,\text{and} \quad \text{for all} \quad\text{$UAV_i$}
\end{equation}
Where the instant all the UAV's identifiers in the network discovered by $UAV_i$ is given by	
\begin{equation}
k^*=min \{k| \quad Id_i(k) = Id_i(k-1)\}
\end{equation}

Then, number of UAVs in the network is given as

\begin{equation}
n=\lvert Id_i(k^*)\rvert
\end{equation}

\end{prop}

\paragraph{proof}
$ $\newline
Note that the identifiers $Id_i(k)$ contains identifiers of $j$ UAVs at $k$ or less hops from $UAV_i$. And $[D^k]_{ij}$ is positive if and only if there exists a direct path of length $k$ from the UAV $i$ to $j$ \cite{bullo2009distributed}. If the set $Id_i$ stop changing, that means all node $j$ at $k$ or less hops are included in the identifier set and as $\mathcal{G}$ is connected, all UAVs are reached with a maximum of $k$ hops from $UAV_i$. Therefore, all UAVs in the network are covered by the paths from $i^{th}$ UAV and the identifiers $Id_i(k)$ has all the identifires of all the network UAVs at the $k^*$ instant. It follows  $n=\lvert Id_i(k^*)\rvert$

\subsection{Estimates of Algebraic Connectivity $(\lambda_2)$}
With all the background regarding power iteration, here we state an algorithm for distributively estimating the algebraic connectivity of the communication graph Laplacian.
\begin{algorithm}[h]
\caption{ Connectivity Estimation Distributively }\label{euclid}
\label{alg:conne_est}
\begin{enumerate}
\item At every step $k$, each UAV  $i$  computes $P(k)$ locally
\begin{equation}\label{eq:D_barUpndateRuleAlg}
	P_i(k)=\sum_{j\in Id_i(k)}  \lvert  [\hat{D}^k]_{ij}-\frac{1}{n} \rvert +\frac{n- \lvert Id_i(k)\rvert }{n}
\end{equation}
\item Run max-consensus to find
\begin{equation}\label{eq:P_max_consensus}
	\lVert P^k \rVert_\infty=\max\limits_{j \in \mathcal{V} } P_j(k)
\end{equation}
\item The algebraic connectivity $\hat{\lambda}_2$ estimated by each UAV $i$ is given by
\begin{equation}\label{eq:lambda_Estimate}
	\hat{\lambda}_{2i}=\frac{(1-\hat{\rho}_i(k))}{\alpha} \quad \text{where}\quad \hat{\rho}_i(k)=(\lVert P^k \rVert_\infty)^{1/k}
\end{equation}
\end{enumerate}
\end{algorithm}

\begin{theorem}
Each UAV (node) $i$ executes Algorithm-\ref{alg:conne_est} with $\mathcal{G}$ connected. As $k \to \infty$, the connectivity eigenvalue estimates, $\hat{\lambda}_{2i}$ converges to the Laplacian algebraic connectivity ${\lambda_2}(\mathcal{L})$ for $\forall \in \mathcal{V}$.
\begin{equation}
\lim_{k \to \infty} \hat{\lambda}_{2i}(k)=\hat{\lambda}_2(\mathcal{L})
\end{equation}
\end{theorem}

\paragraph{proof}
$ $\newline
For $\forall i \in \mathcal{V}$, the variable $[\hat{D}^k]_{ij}$ are equal to $[D^k]_{ij}$ for $ j \in Id_i(k)$, where $[D^k]_{ij}=0$ for $ j \notin Id_i(k)$, where from (\ref{eq:P_power_k}), $P^k=D^k-\textbf{11}^T/n$, gives the following
\begin{equation}
\begin{split}
	[P^k]_{ij}&=[\hat{D}^k]_{ij}-1/n \quad \text{for} \quad j \in Id_i(k)\\
	[P^k]_{ij}&=-1/n \quad \quad \quad\text{for} \quad j \notin Id_i(k), \quad \text{for} \quad \forall i \in \mathcal{V}, k\geq 1
\end{split}
\end{equation}

When $\mathcal{G}$ is connected, $\lVert P^k \rVert_\infty$ in (\ref{eq:P_max_consensus}) gives the maximum absolute row sum of $P^k$ and the spectral estimate of $P^k$, $\hat{\rho}_i(k)$ in (\ref{eq:lambda_Estimate}) is 			
\begin{equation}
\hat{\rho}_i(k)= \lVert P^k \rVert^{1/k}_\infty \quad \text{for} \quad \forall i \in {1,...,n} \quad k\geq 1
\end{equation}
For induced $\infty$-norm, we have,
\begin{equation}
\hat{\rho}_i(k)= \lim_{k \to \infty} \lVert P^k \rVert^{1/k}_\infty
\end{equation}
Hence
\begin{equation}
\lim_{k \to \infty} \hat{\lambda}_i(k)=\lim_{k \to \infty} \frac{(1-\hat{\rho}_i(k))}{\alpha}=\hat{\lambda}_2(\mathcal{L})
\end{equation}

\subsection{Estimates of Fiedler vector, $v_2$}

Fiedler vector, the eigenvector corresponding the algebraic connectivity, can be computed distributively by noticing some facts and relationship regarding the non-deflated matrix and the Laplacian matrix eigenvectors.
\begin{enumerate}
\item  In (\ref{eq:defD}),  $D=\textbf{I}-\alpha L_{\mathcal{G}}$, notice that all the eigenvectors of $D$ and $L_{\mathcal{G}}$ are the same \cite{olfati2007consensus}.
\item Eigenvectors do not change during matrix power operation.
Let $v$ be an eigenvector of matrix $D$, hence $Dx=\lambda x$, then
\begin{equation}
\begin{split}
	D^kx&=D^{k-1} (Dx)=D^{k-1} (\lambda x)=\lambda D^{k-1} x \\
	&=\lambda^k x
\end{split}
\end{equation}
\end{enumerate}

Notice that from Proposition \ref{Prp: max step to identify n}, at step $k^*$, UAV $i$ has variables $[\hat{D}]_{ij}$ for all the agents $j \in Id_i(k^*)$ that are at $k^*$ or less hops from $i$. For the remaining agents $j \notin Id_i(k^*)$, UAV $i$ does not have variable $[\hat{D}]_{ij}$, but as all the UAVs are already in $k^*$-hop reach and explored, then their value of $[\hat{D}]_{ij}$ of $j \notin Id_i(k^*)$ is zero then.

\begin{algorithm}[h]
\caption{Estimates of Fiedler Vector }\label{ alg:Fiedler vector euclid}

\begin{enumerate}
\item At $k^*$, each UAV $i$ initialized a matrix $\tilde{D}^i$ of $n \times n$
\begin{equation}
	\tilde{D}^i (0)=\textbf{0}
\end{equation}
\item Run consensus so that each UAV $i$  update its private non-deflated matrix $\tilde{D}^i$
\begin{equation}
	[\tilde{D}^i]_{ij}=[\hat{D}]^{k^*}_{ij}
\end{equation}
After consensus, get $\tilde{D}^i$ of each UAV $i$.
\item Each UAV $i$ compute the eigenvalue and eigenvector of $\tilde{D}^i$ in (2) which is the same for all $i \in \mathcal{V}$.
\begin{equation}
	[\tilde{\mathbf{v}}^i \quad \tilde{\boldmath{\lambda}}^i]=eignvalue(\tilde{D}^i)
\end{equation}
where
\begin{enumerate}
	\item
	$
	\tilde{\boldmath{\lambda}}^i=\{\tilde{\lambda}^{k^*}_1,\tilde{\lambda}^{k^*}_2...,\tilde{\lambda}^{k^*}_n\},\quad \text{ and }  \tilde{\lambda}^{k^*}_1 <\tilde{\lambda}^{k^*}_2< ... <\tilde{\lambda}^{k^*}_n. \text{ (all to the power of } k^* )
	$
	\item \text{the corresponding eigenvectors,}\\ $\tilde{\mathbf{v}}^i=\{\tilde{v}_1,\tilde{v}_2,...,\tilde{v}_n\}$
\end{enumerate}

\item The Fiedler vector of the Laplacian is $v_2=\tilde{v}_2$
\end{enumerate}
\textit{\textbf{Note}}: $D=\textbf{I}-\alpha L_{\mathcal{G}}$, all the eigenvector of $D$ and $L_{\mathcal{G}}$ are the same. Further more, power operation of on a matrix does not change the matrix eigenvectors. Hence, eigenvector of $D^k$ and $L_{\mathcal{G}}$ are the same.
\end{algorithm}

\section{Resilient UAV Formation Control with Mis-behaving UAVs}

For a resilient formation flight of UAVs, here a two stage controller is proposed. The UAV first need to guarantee the $(2F+1)$-robustness of the underlying communication network before pursuing a formation flight with W-MSR algorithm. This is to guarantee the normal nodes to achieve consensus on the formation flight and route despite the presence of up to $F$ misbehaving UAVs in the network.

\subsection{Achieving $(2F+1)$-robustness}
The UAVs first try to achieve $(2F+1)$-robustness by increasing the algebraic connectivity of their communication graph. The controller tries to achieve this by applying

\begin{equation}\label{eq:lambda_2 gradient control01}
\textbf{u}_i=\frac{\partial \lambda_2(x)}{\partial x_{i,\alpha}}={ \nabla } _i \lambda_2
\end{equation}

where,  ${ \nabla }  \lambda_2 \in \mathbf{R}^{dN}$ ( with $d$-dimension, $N$- number of UAVs or nodes), is determined from equation \ref{eq:lambda_2 gradient}, $\textbf{u}_i \in \mathbf{R}^d $  and ${ \nabla } _i \lambda_2$ is $d$-dimensional gradient that steers UAV $i$.

The knowldge of the derivative of $\lambda_2(x)$ gives local directions to move to improve or at least to maintain connectivity of the network as measured by $\lambda_2$. Hence, the controller in (\ref{eq:lambda_2 gradient control01}) based on the $d$-dimensional gradient, ${ \nabla } _i \lambda_2$, drives the $i^{th}$ UAV to be more connected with the rest of the UAVs and subsequently achieve resilient connectivity threshold.

\begin{prop}
For system of UAVs in the range of communication radius  $\lVert x_i-x_j \lVert  \leq R$, the controller in (\ref{eq:lambda_2 gradient control01}) will steer the system to a resilient state.
\end{prop}

\paragraph{proof}
$ $\newline
As long as the UAVs are in the communication range $\lVert x_i-x_j \lVert  \leq R$, and as long as $\lambda_2$ is less than the threshold $\lambda_2 \leq 4F$, the controller will be applied to all $UAV_i$ $i \in \mathcal{N}_j$, hence resilient state will be achieved.

\subsection{Resilient Formation}
The UAVs once in a resilient state, the formation control will be applied to drive them into a resilient formation flight.

\begin{equation}\label{eq:lambda_2 gradient and formation cont}
\textbf{u}_i=  \varphi{ \nabla } _i \lambda_2 +\textbf{u}{_{formation}}_i
\end{equation}
where $\textbf{u}{_{formation}}_i $ is given in equation \ref{eq:control enput W-MSR} which is for each dimension of the UAV motion, $i^{th}$ UAVs control input for formation flight of the UAV system.
\begin{prop} \label{pro: resilient control}
If the UAVs driven to resilient state following controller in (\ref{eq:lambda_2 gradient control01}) and then followed by a controller in (\ref{eq:lambda_2 gradient and formation cont}), they will not fall back to vulnerable state. In addition, if there is no more than $F$ miss-behaving UAVs, the UAVs converge asymptotically to the designated formation.
\end{prop}

\paragraph{proof}
$ $\newline
\begin{enumerate}
\item The condition to fall back to a non-resilient (volunerable) state is that when $\lambda_2$ goes down below the stated resilient threshold. It suffice to show that
\begin{equation}
	\frac{d \lambda_2}{dt}\geq 0, \quad \forall t
\end{equation}

The derivative can be written as using chain rule
\begin{equation}
	\frac{d \lambda_2}{dt}=\sum\limits_{i=1}^{N} \frac{\partial  \lambda_2}{\partial x_i } \frac{\partial x_i}{\partial t }=\sum\limits_{i=1}^{N} \frac{\partial  \lambda_2}{\partial x_i } (A_{i}x_i+B_{i} u_i)
\end{equation}
Then, as long as
\begin{equation} \label{eq:lambda derivative inequality}
	(\frac{\partial  \lambda_2}{\partial x_i })^T (A_{i}x_i+B_{i}(\varphi{ \nabla } _i \lambda_2 + \textbf{u}{_{formation}}_i )  \geq 0,\quad \forall i, \quad \forall t
\end{equation}
the UAVs will never reenter to the non-resilient state. The value of $\varphi$ is taken to make sure the inequality (\ref{eq:lambda derivative inequality}) holds for the system of UAVs.  Solving for $\varphi$,
\begin{equation} \label{eq:gama}
	\varphi \geq \frac{{ \nabla } _i \lambda_2  ^T(A_{i}x_i+B_{i} \textbf{u}{_{formation}}_i )}{{ \nabla } _i \lambda_2  ^T B_i{ \nabla } _i \lambda_2   }
\end{equation}
where the above equation (\ref{eq:gama}) can be solved by each $i^{th}$ UAV at each time step, making sure the system does not return to the non-resilient (vulnerable) state.
\item The network of UAVs with $\lambda_2>4F$ in the presence of $F$ non cooperating UAVs will guarantee $(2F+1)$ robustness. Therefore, W-MSR algorithm derives the network of UAVs  to a consensus, in this case a formation flight.
\end{enumerate}

\noindent
Further more, once the UAVs entered a resilient state, applying only the formation control

\begin{equation}\label{eq:formation control input }
\textbf{u}_i=  \textbf{u}{_{formation}}_i
\end{equation}
keeps the system in a resilient stat,.i.e, it will not slip back to the vulnerable states.

\begin{prop}
A network of UAVs once in a resilient stat will remain in resilient state under the formation control for a misbehaving agents less than F.
\end{prop}

\paragraph{proof}
$ $\newline
In the previous proposition \ref{pro: resilient control},  we showed that the control scale of $ \varphi$ to guarantee $\frac{d \lambda_2}{dt} \geq 0 $, and to make sure the controller keeps the system of networked UAVs under $(2F+1)$ robustness consequently a resilient one. But here it is enough to show $\frac{d \lambda_2}{dt} $ converge exponentially to zero in finite time in a resilient state as the UAVs reach consensus in the formation flight.
\begin{equation}
\begin{split}
	\frac{d \lambda_2}{dt}&=\sum\limits_{i=1}^{N} \frac{\partial  \lambda_2}{\partial x_i } \frac{\partial x_i}{\partial t }\\
	&=\sum\limits_{i=1}^{N} \frac{\partial  \lambda_2}{\partial x_i } (A_{i}x_i+B_{i}\textbf{u}{_{formation}}_i)
\end{split}
\end{equation}
\noindent
Taking the details in the sumation

\begin{equation}
\frac{d \lambda_2}{dt}=\sum\limits_{n=1}^{N} \frac{\partial  \lambda_2}{\partial x_i } (A_{i}x_i+B_{i}\textbf{u}{_{formation}}_i )
\end{equation}
From \ref{eq:lambda_2 gradient}, we have
\begin{equation}\label{eq:lambda_2 gradient_second equation }
\frac{\partial \lambda_2(x)}{\partial x_{i,\alpha}}= Trace \left \{ \left [\frac{v_2 v_2^T}{v_2^T v_2}\right ]^T \left [\frac{\partial L_{\mathcal{G}}(x)}{\partial x_{i,\alpha}} \right ]   \right \}
\end{equation}

Where the Laplacian matrix entries are given by

\begin{equation} \label{eq:lap_in_short}
[L_{\mathcal{G}}]_{ij}=
\begin{cases}
	-f_{ij}(x),\quad i\neq j\\
	\sum_{l=1,l\neq i}^{N} f_{il}(x) \quad i=j
\end{cases}
\end{equation}
Looking into the Laplacian (\ref{eq:lap_in_short}) derivative, $\left [\frac{\partial L_{\mathcal{G}}(x)}{\partial x_{i,\alpha}} \right ]$, with respect to the spatial dimension of $\alpha \in 1...d$, the off diagonal entries are non zero for UAVs in the communication range.

\noindent
Since
\begin{equation}
\frac{\partial f_{ij}(x)}{\partial x_{i,\alpha}}= -\frac{\partial f_{ij}(x)}{\partial x_{j,\alpha}}
\end{equation}
the off-diagonal entries of $ \sum_{k}^{N} \frac{\partial L_{\mathcal{G}}}{\partial x_{k,\alpha}},$ where, $ k \in \{ i,j \}$, and for each spatial dimension of $\alpha \in 1...d$ is given as

\begin{equation}
\begin{split}
	\sum_{k}^{N} -\frac{\partial f_{ij}(x)}{\partial x_{k,\alpha}}&= -\frac{\partial f_{ij}(x)}{\partial x_{i,\alpha}}(A_{i}x_{i,\alpha}+B_{i}\textbf{u}{_{formation}}_i)-\frac{\partial f_{ij}(x)}{\partial x_{j,\alpha}}(A_{i}x_{j,\alpha}+B_{i}\textbf{u}{_{formation}}_i)\\
	&=-\frac{\partial f_{ij}(x)}{\partial x_{i,\alpha}}A_{i}(x_{j,\alpha}-x_{i,\alpha})
\end{split}
\end{equation}

And the diagonal entries are
\begin{flalign*}
\sum_{k}^{N} \sum_{j}^{N} \frac{\partial f_{ij}(x)}{\partial x_{k,\alpha}} (A_{i}x_{k,\alpha}+B_{i}\textbf{u}{_{formation}}_i)=&\sum_{j}^{N} \sum_{k}^{N} \frac{\partial f_{ij}(x)}{\partial x_{k,\alpha}}(A_{i}x_{k,\alpha}+B_{i}\textbf{u}{_{formation}}_i)&\\
&=\sum_{j}^{N} \bigg(\frac{\partial f_{ij}(x)}{\partial x_{i,\alpha}}(A_{i}x_{i,\alpha}+B_{i}\textbf{u}{_{formation}}_i)\\
&+ \frac{\partial f_{ij}(x)}{\partial x_{j,\alpha}}(A_{i}x_{j,\alpha}+B_{i}\textbf{u}{_{formation}}_i\bigg)&\\
&=\sum_{j}^{N}\frac{\partial f_{ij}(x)}{\partial x_{i,\alpha}}A_{i}(x_{j,\alpha}-x_{i,\alpha} )
\end{flalign*}
The value $| x_{j,\alpha}-x_{i,\alpha} | \longrightarrow X_d$, therefore, it approaches asymptotically to $\mid X_d \mid$ of the separation distance in the spatial dimension and approaches asymptotically to zero in its heading velocities as the UAVs are running the consensus algorithm. As the rate of convergence of the W-MSR algorithm is exponential, hence the $| x_{j,\alpha}-x_{i,\alpha} | \longrightarrow X_d$ also converges exponentially.
\\In a formation set up, once the UAVs are in position within a communication range $\lVert x_i-x_j \lVert < R$, then $\lVert x_i-x_j \lVert \longrightarrow $ fixed, as the distance between UAVs do not change, therefore the $L_{\mathcal{G}}(x)$ is constant as
\begin{equation}
\frac{\partial L_{\mathcal{G}}(x)}{\partial x_{i,\alpha}} =0
\end{equation}
Hence,
\begin{equation}\label{eq:lambda_2 gradient_second equation2 }
\begin{split}
	\sum_{i} \frac{\partial \lambda_2(x)}{\partial x_{i,\alpha}}&= \sum_{i} Trace \left \{ \left [\frac{v_2 v_2^T}{v_2^T v_2}\right ]^T \left [\frac{\partial L_{\mathcal{G}}(x)}{\partial x_{i,\alpha}} \right ]   \right \}\\
	&=Trace \left \{ \left [\frac{v_2 v_2^T}{v_2^T v_2}\right ]^T \sum_{i} \left [\frac{\partial L_{\mathcal{G}}(x)}{\partial x_{i,\alpha}} \right ]   \right \}\\
	&=0
\end{split}
\end{equation}

that is, each element of the gradient is zero.

Consequently,
\begin{equation}
\frac{d \lambda_2}{dt}=0
\end{equation}

\section{Simulation Results}
In the simulations, different scenarios are considered in which a team of UAVs  tries to achieve a resilient formation flight in the presence of misbehaving UAVs starting from a random location and assumed to be on a level flight. Each UAV is within a communication radius of another UAV, making it a connected graph.
\subsection{Misbehaving UAVS} \label{misbehaving UAVs selection}
The misbehaving UAV(s) could be picked randomly out of the $20$ UAVs to show either failure or being a victim of a cyber attack. But in an intelligent cyber attack case, the UAV node in the communication graph with the maximum connection or signal strength picked  for most adverse potential to influence the other cooperative UAVs.

Here we consider $F=\mid\mathcal{N}_a\mid=2$ misbehaving UAVs with a different influencing factors with: a) the largest communication link with its immediate neighbors b) the strongest link signal strength, i.e., $max \{  [L_{\mathcal{G}}(x)]_{ij} \}  $ for all $i=j$.

Threat Model: The UAVs use a wireless broadcasting method for their communication hence a threat model of `Malicious UAV'  will characterize it well. The threat in a formation flight, where a misbehaving UAV $l\in \mathcal{N}_a$ is a malicious one if it does not follow the prescribed update rule. It send out the same value for all its neighbors i.e.,
\begin{equation}
x_{(l \rightarrow i)}(t)\equiv x_{(l\rightarrow j)}(t),\quad \forall i, j \in \mathcal{N}_l.
\end{equation}

This will help to model the misbehaving UAVs being under an intelligent cyber attack but also to the least, to act as a random failure of UAVs as well.

\subsection{Regular Consensus Algorithm } \label{Regualr consesus without falulty UAV}
\begin{figure}[htb]
\subfigure[Intial Random Graph ]{\includegraphics[width = .5\columnwidth]{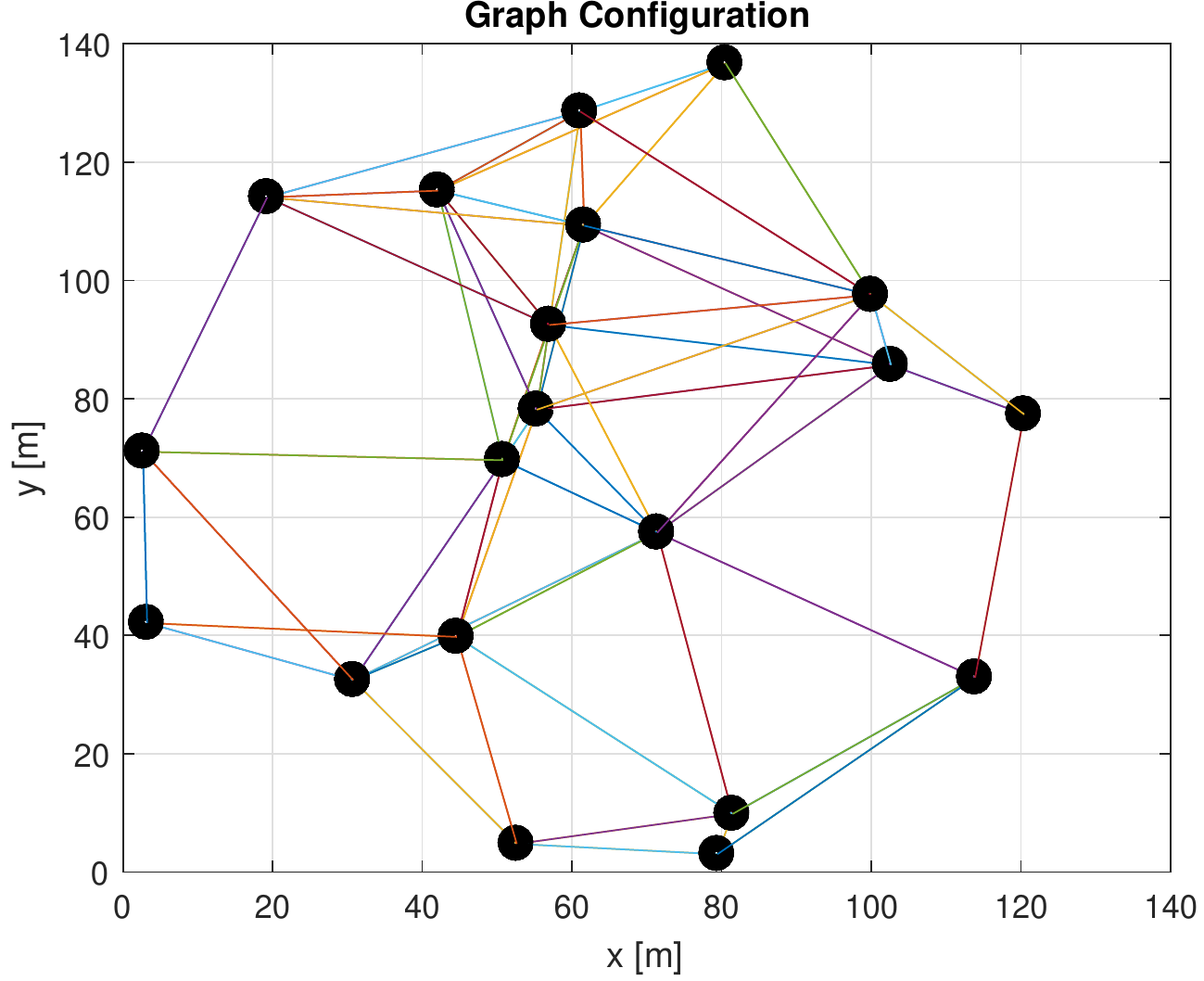}\label{ReguConsIntial graph}}
\subfigure[Formation flight]{\includegraphics[width = .5\columnwidth]{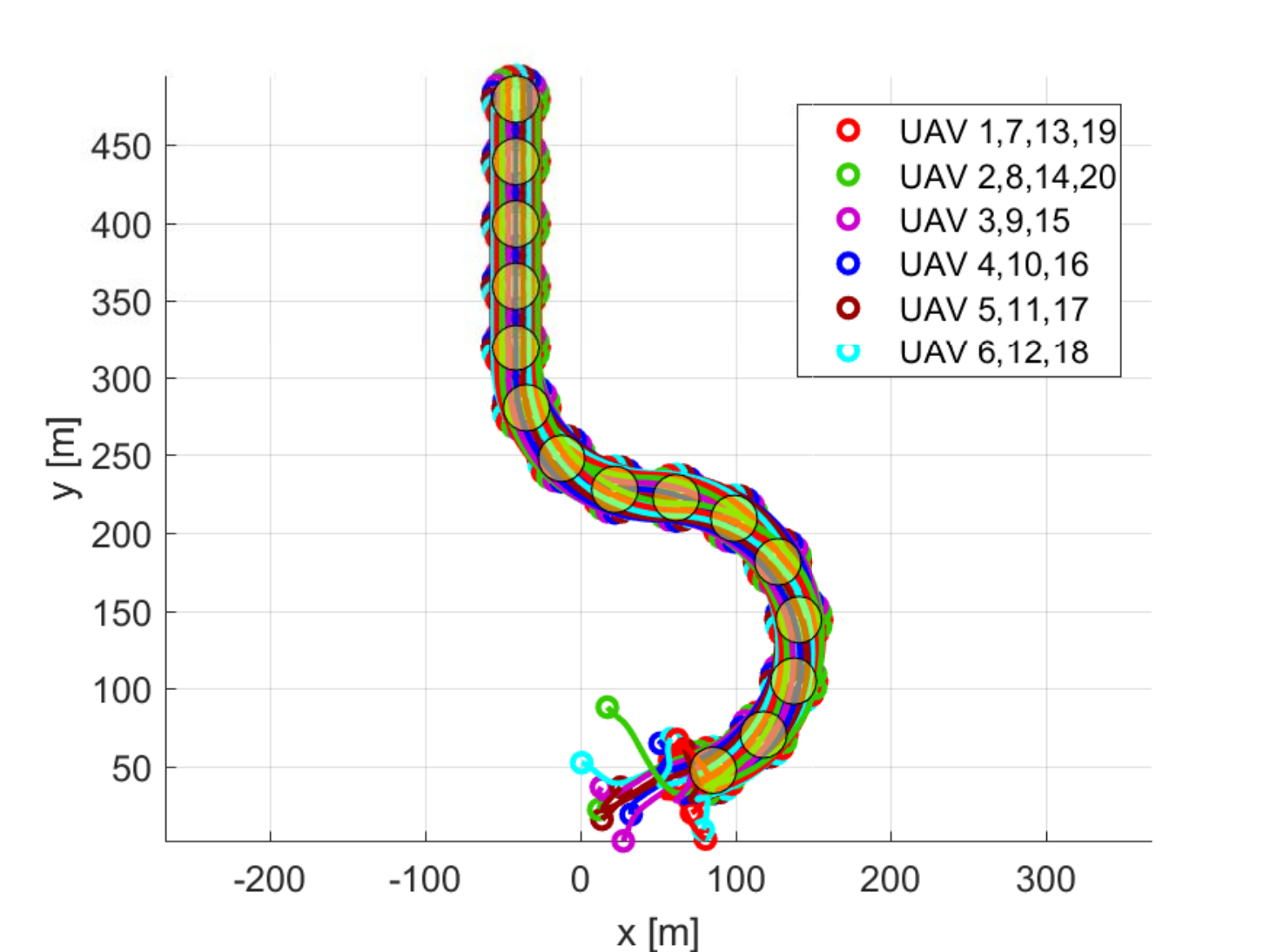}\label{ReguConsFormation}}
\caption{(a) Initial communication graph for 20 UAVs  (b) UAVs under 20-gon formation fly.}
\label{ReguCongraph}
\end{figure}

\begin{figure}[h]
\subfigure[Final communication graph ]{\includegraphics[width = .5\columnwidth]{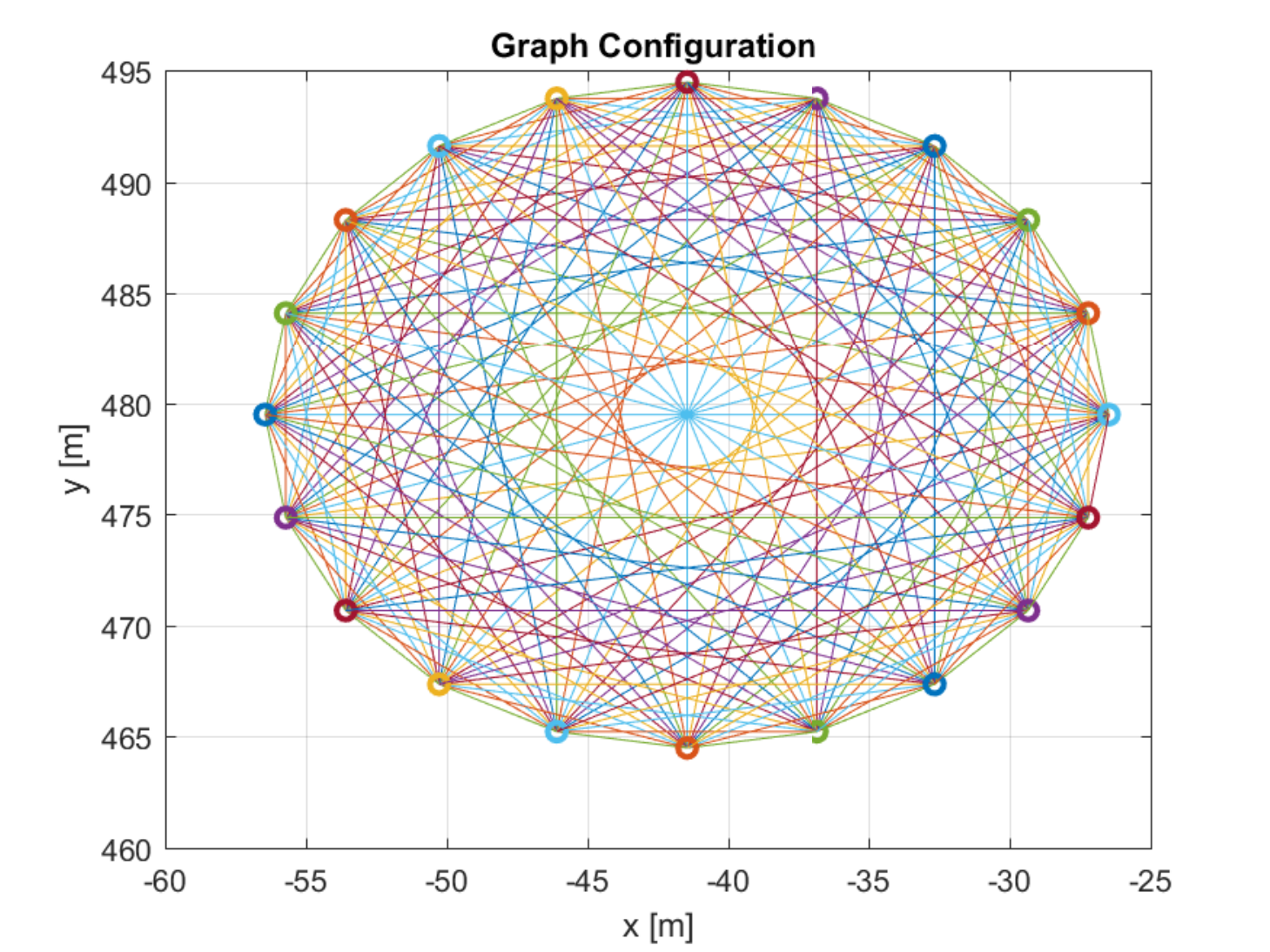}\label{ReguConsFinal graph}}
\subfigure[Algebraic connectivity $\lambda_2(t)$  ]{\includegraphics[width = .5\columnwidth]{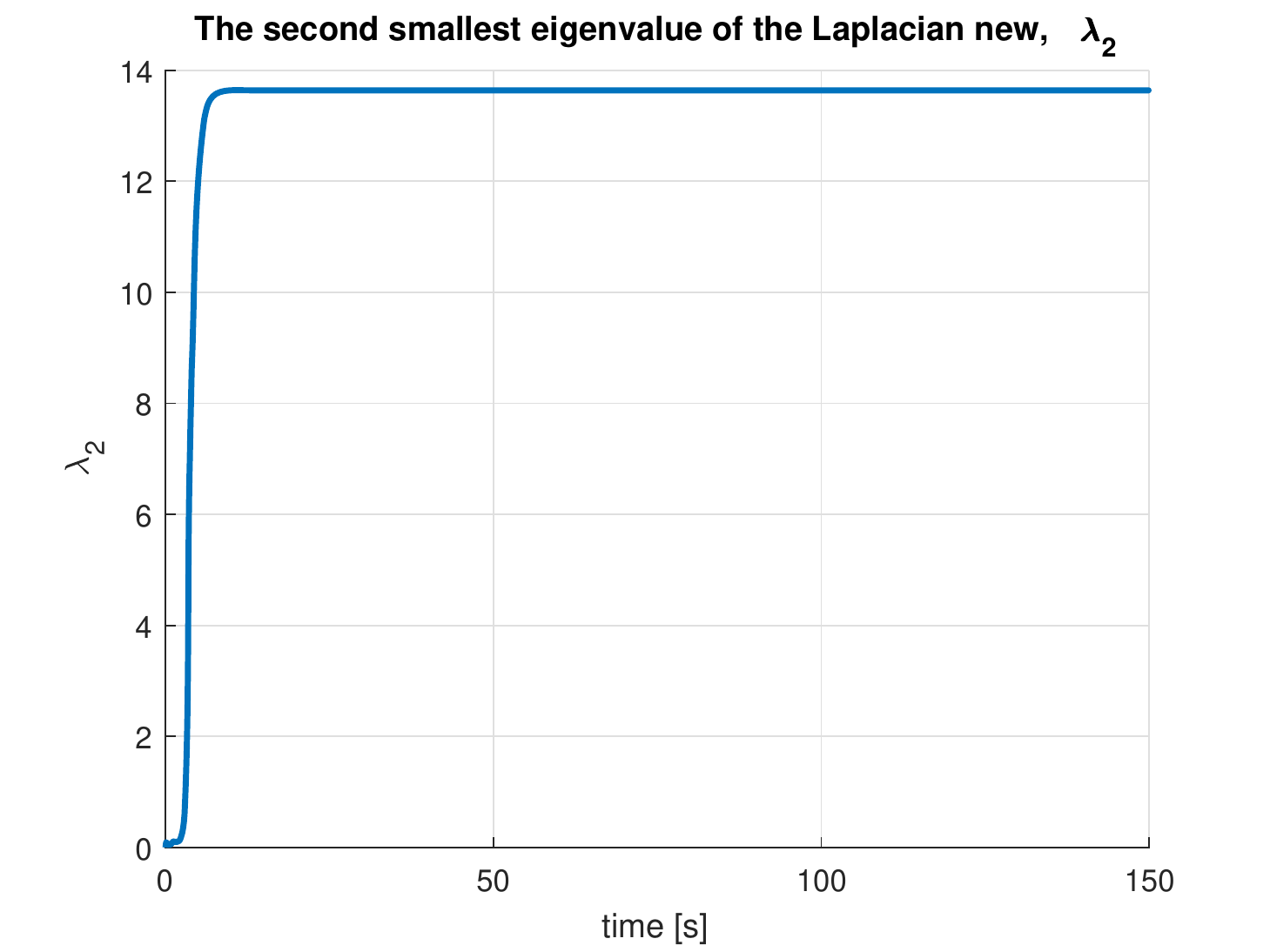}\label{ReguConsEigneVal-2}}
\caption{(a) The 20 UAVs communication graph with different strength (indicated with different color of edges)  (b) the algebraic connectivity $(\lambda_2(t))$ of the UAVs under formation fly.}
\label{ReguConFinalConEign}
\end{figure}

All well behaving $20$ UAVs start from a random location (Fig. \ref{ReguConsIntial graph}) and the connectivity maintenance controller and formation control in  (\ref{eq:lambda_2 gradient and formation cont}) applied to the system of UAVs in level flight. The controller is able to keep the UAVs in the specified 20-gon (icosagon) formation (Fig. \ref{ReguConsFormation} and heading straight in the positive $y-$direction with $[v_x=0 m/s \quad v_y=4m/s]^T$ after flying half a circle. The controller also maintains the algebraic connectivity of the graph Laplacian well above $\lambda_2=13$ (Fig. \ref{ReguConsEigneVal-2}). The final communication graph in Figure \ref{ReguConsFinal graph} shows that the controller managed to bring all the UAVs into communication radius range with certain connectivity level while keeping the formation in 20-gon form.

\begin{figure}[htbp]
\subfigure[x-velocity ]{\includegraphics[width = .5\columnwidth]{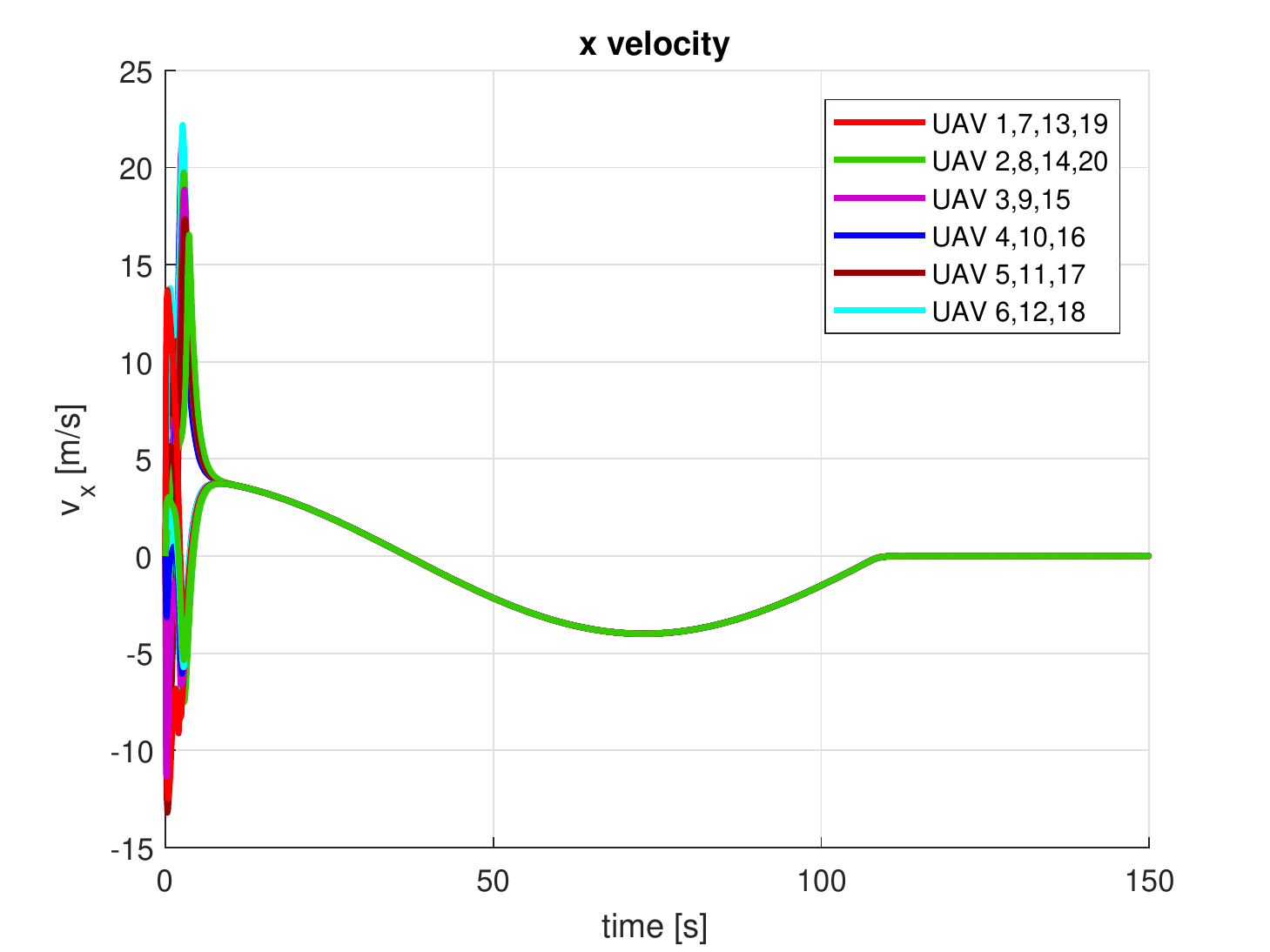}\label{ReguConsX-velo}}
\subfigure[x-position ]{\includegraphics[width = .5\columnwidth]{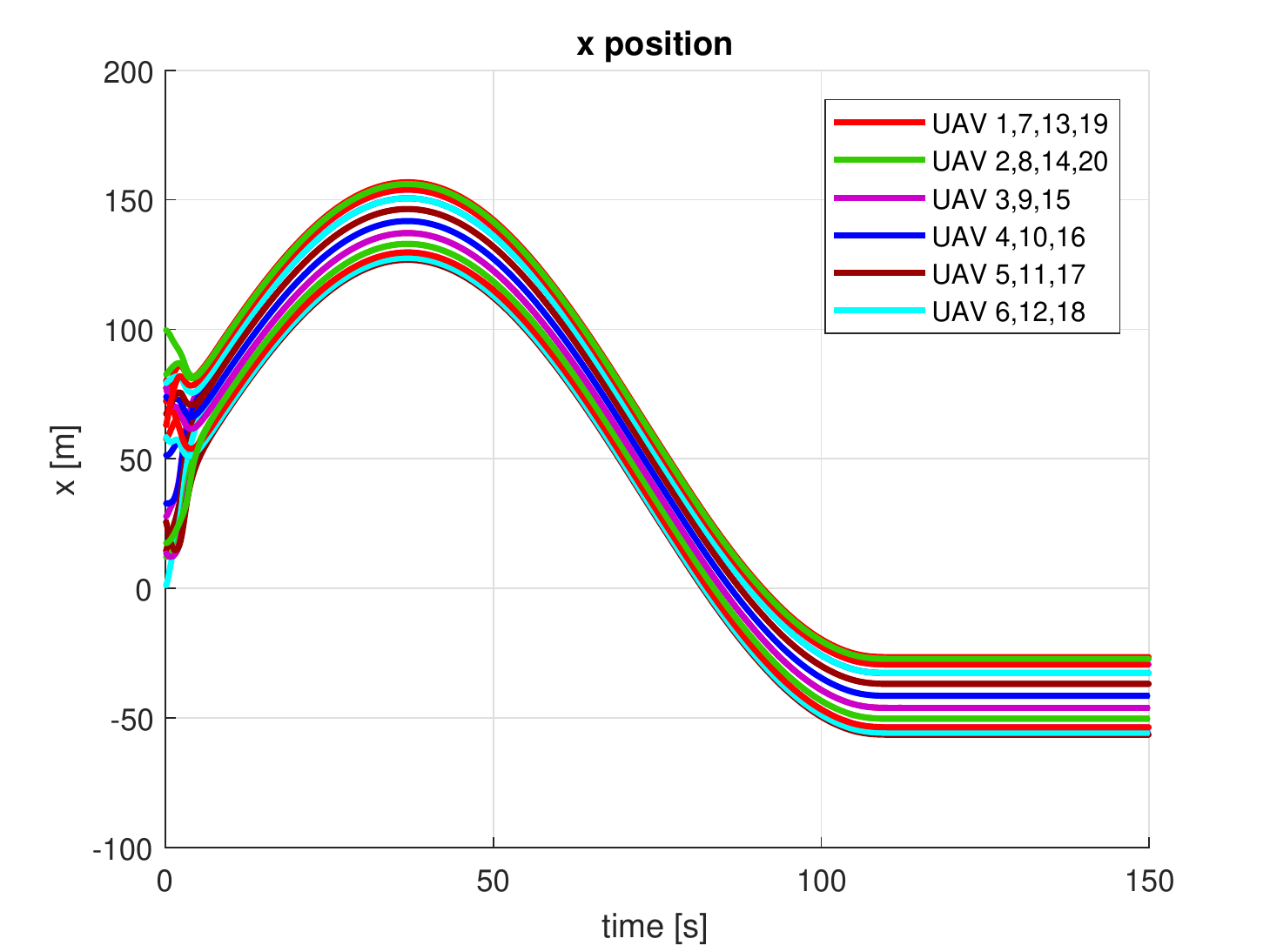}\label{ReguConsX-po}}
\caption{(a) The 20 UAVs x-direction velocities  (b) x-positions of the UAVs under formation fly.}
\label{ReguConX-dim}
\end{figure}

\subsection{Regular Consensus Algorithm with Misbehaving UAVs }
\subsubsection{ Misbehaving UAVs with constant heading}
\begin{figure}[h]
\subfigure[Intial Random Graph ]{\includegraphics[width = .5\columnwidth]{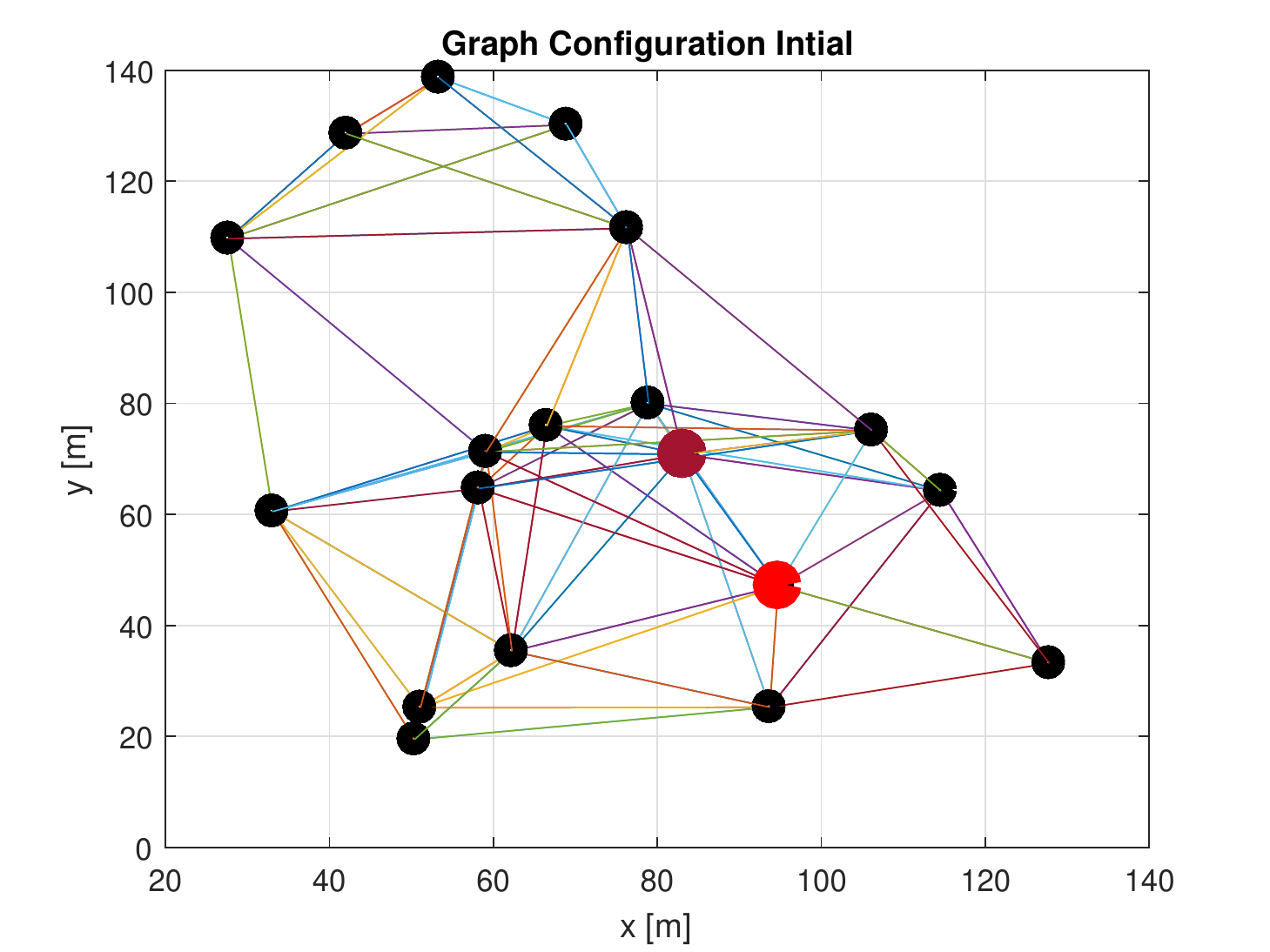}\label{ReguConsIntial graphF}}
\subfigure[Formation flight]{\includegraphics[width = .5\columnwidth]{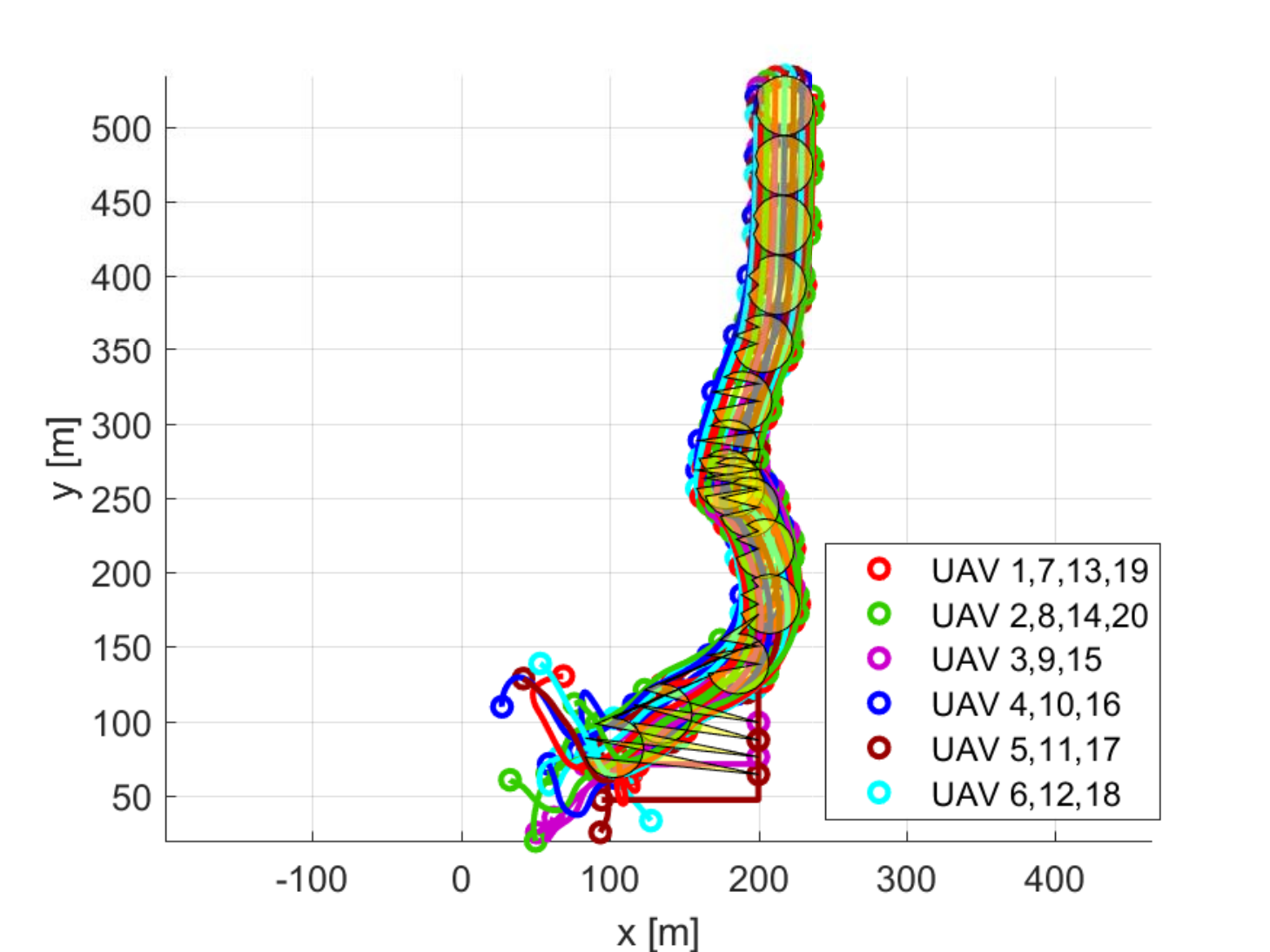}\label{ReguConsFormationF}}
\caption{(a) Initial communication graph with two non-cooperative UAVs, red and dark red colors.  (b) Formation flight of the 20 UAVs but drifted the route towards non-cooperative UAV fixed x-postion, $x=200m$.}
\label{regularCons 1}
\end{figure}

\begin{figure}[h]
\subfigure[x-velocity ]{\includegraphics[width = .5\columnwidth]{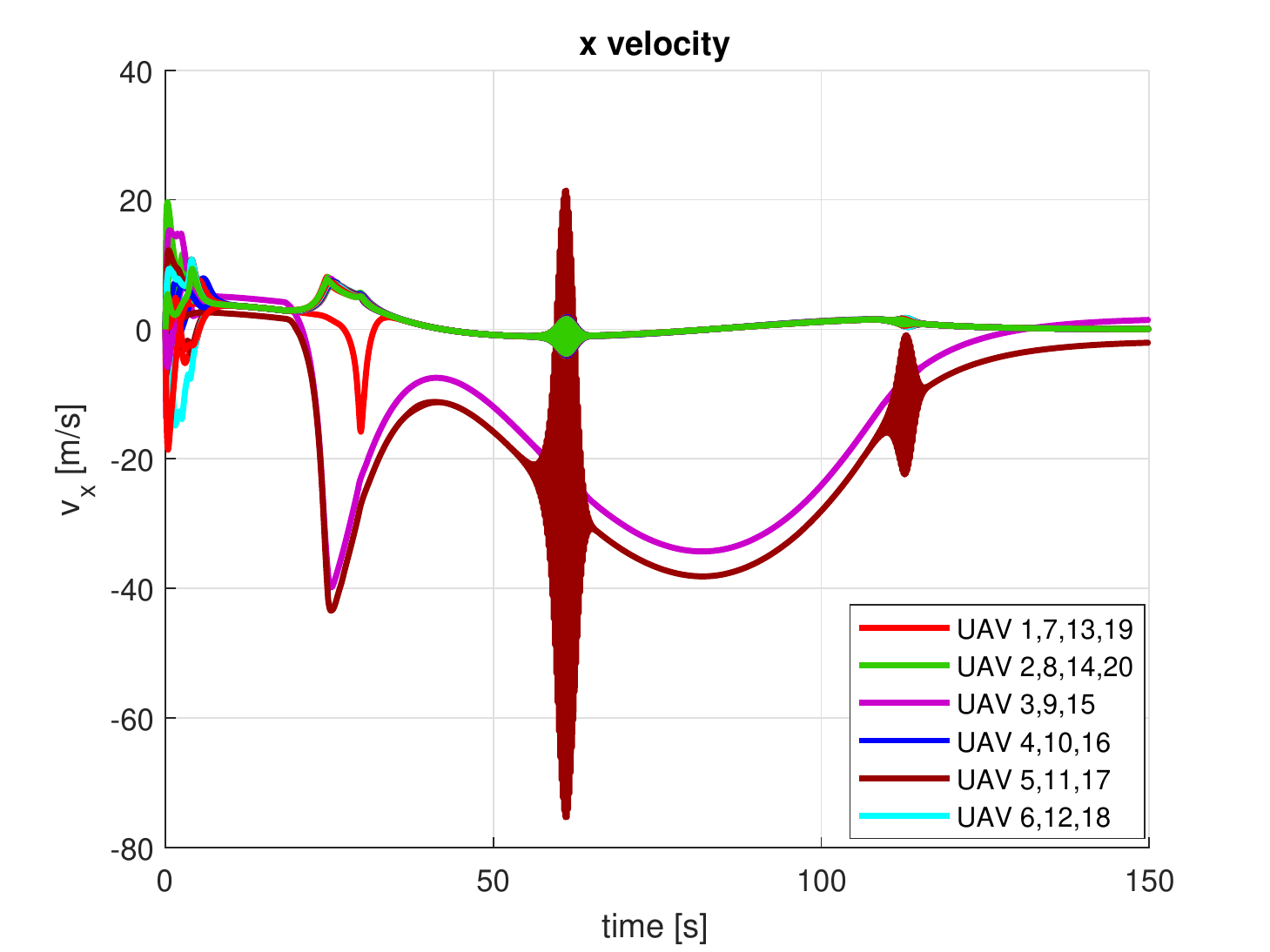}\label{ReguConsx-veloF}}
\subfigure[x-position ]{\includegraphics[width = .5\columnwidth]{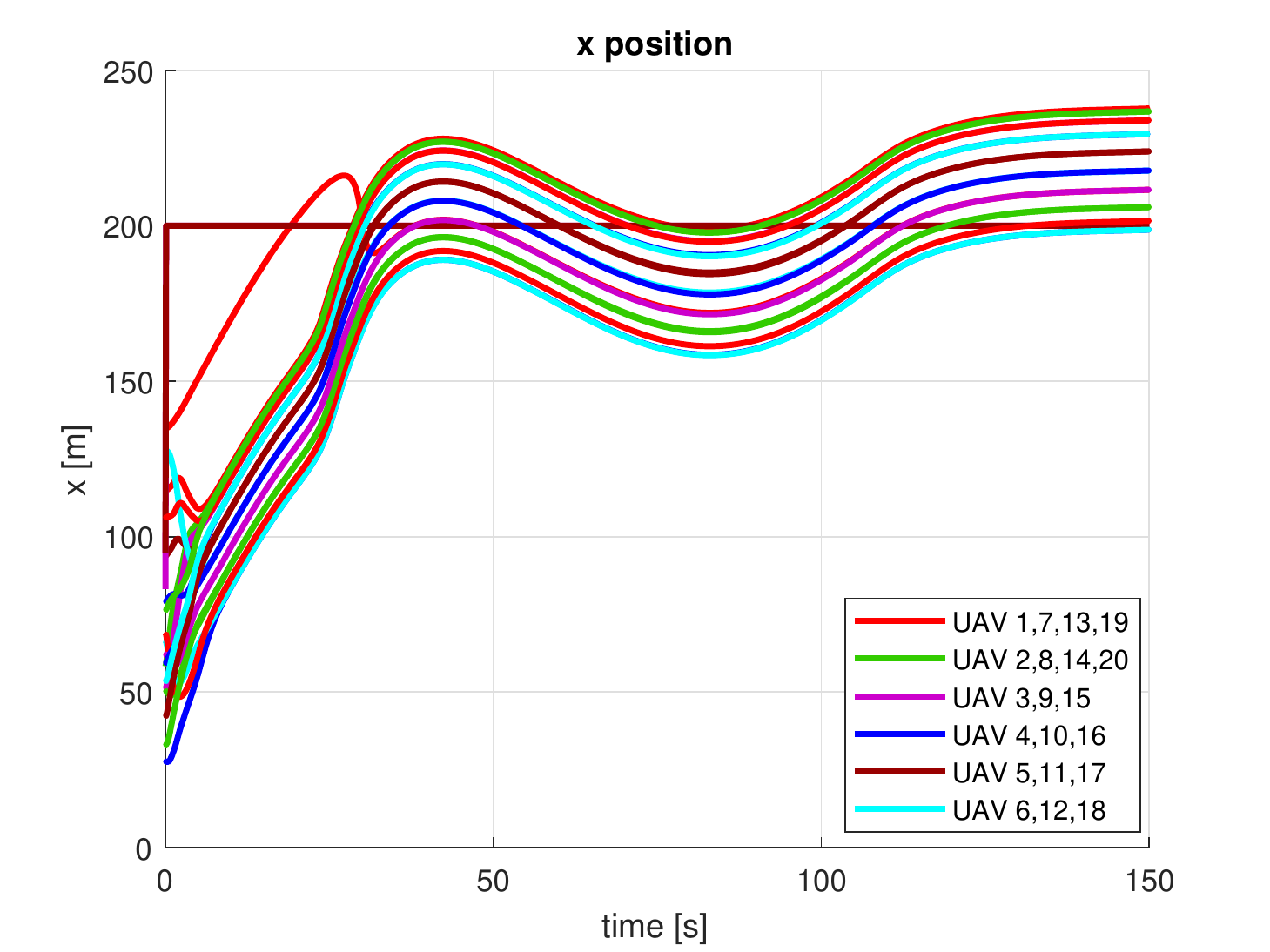}\label{ReguConsReguConsx-poF}}
\caption{(a) x-direction velocities  (b) x-positions of the 20 UAVs under formation flight heavily influence or drifted to the non-cooperative UAVs' x-postion.}
\label{regularCons 1}
\end{figure}

For a randomly generated graph in Figure \ref{ReguConsIntial graphF}, two (i.e., $F=2$)  non-cooperative UAVs are picked. A regular consensus algorithm applied with a connectivity controller to the randomly placed UAVs. The two non-cooperative UAVs maintain a constant x-position at $x=200m$ as seen in Figure  \ref{ReguConsReguConsx-poF}.

Every UAV updates its velocity and position  based on current consensus value $y_i$. In Figure \ref{ReguConsReguConsx-poF}, the two non-cooperative UAVs are sharing a constant value all the time which force the average consensus value to converge around their constant value, $x_{p_i}=200, \forall i\in \mathcal{N}_a $, which ultimately changed the formation flight route as seen in Figure \ref{ReguConsFormationF}. Note that although consensus can be reached, the non-cooperative or malicious UAVs have the potential to drive the consensus process to any value outside of the initial values of the well-behaving UAVs by choosing its initial value.
\begin{figure}[h]
\subfigure[Final communication Graph ]{\includegraphics[width = .5\columnwidth]{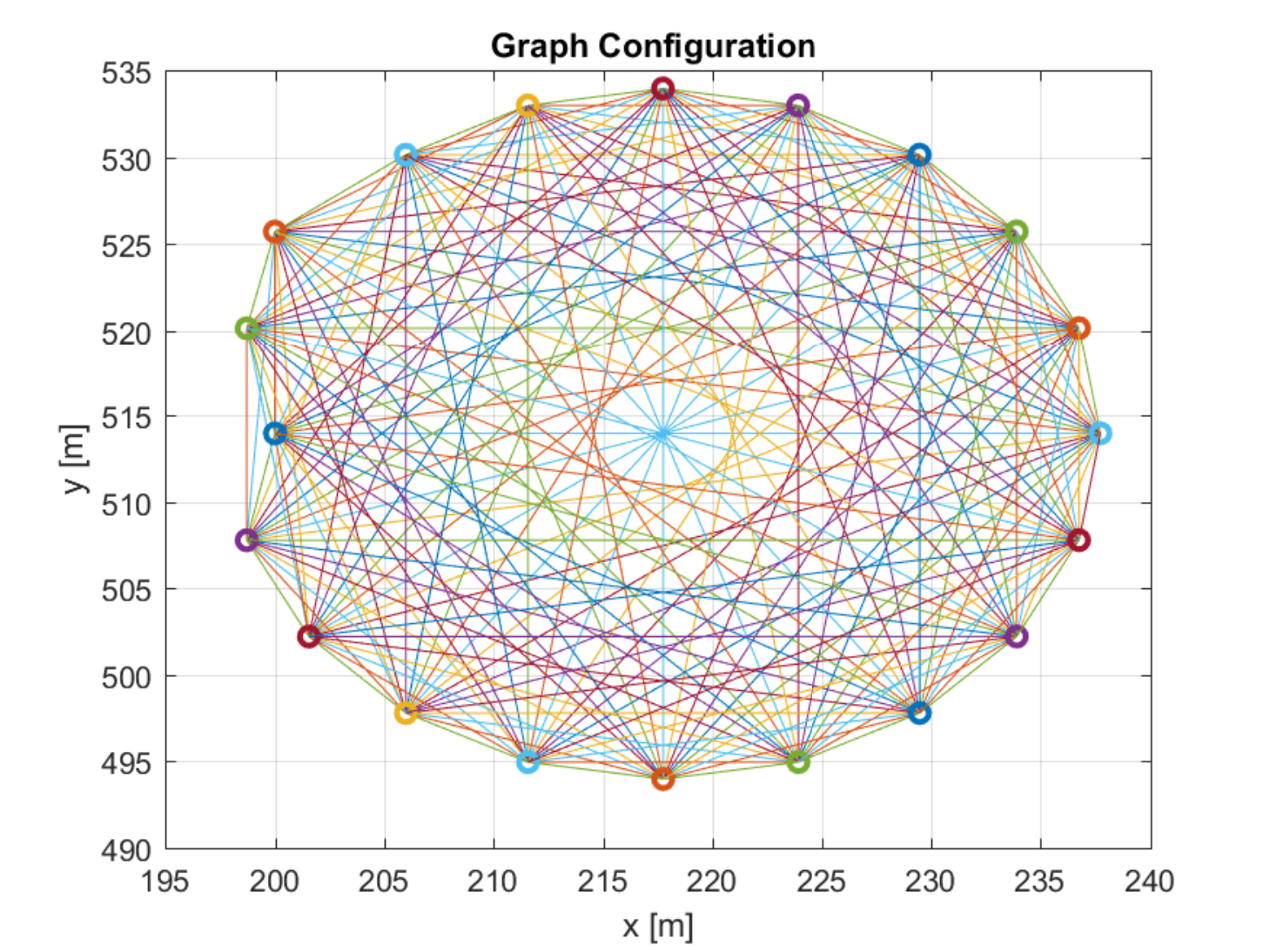}\label{ReguConsfinal graphF}}
\subfigure[Algebric connectivity $\lambda_2(t)$  ]{\includegraphics[width = .5\columnwidth]{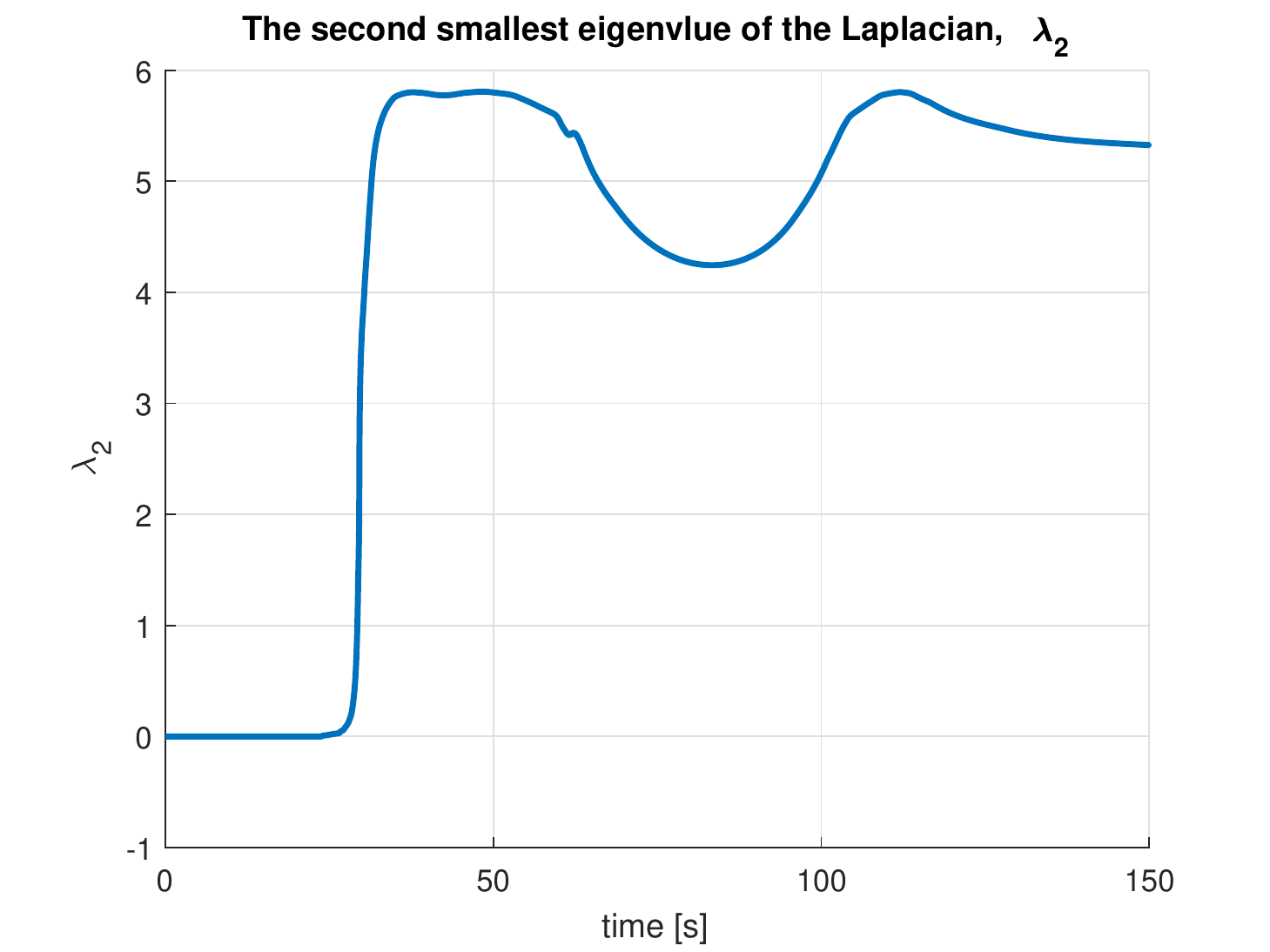}\label{ReguConseigneVal-2F}}
\caption{(a) Final communication graph of the formation flying UAVs under regualr consensus algorithm.  (b) Algebraic connectivity of the system of UAVs under formation flying.}
\label{regularCons 1}
\end{figure}

\subsection{W-MSR Algorithm with Misbehaving UAVs }
\subsubsection{Misbehaving UAVs with Constant Heading}
\begin{figure}[H]
\subfigure[Intial Random Graph ]{\includegraphics[width = .5\columnwidth]{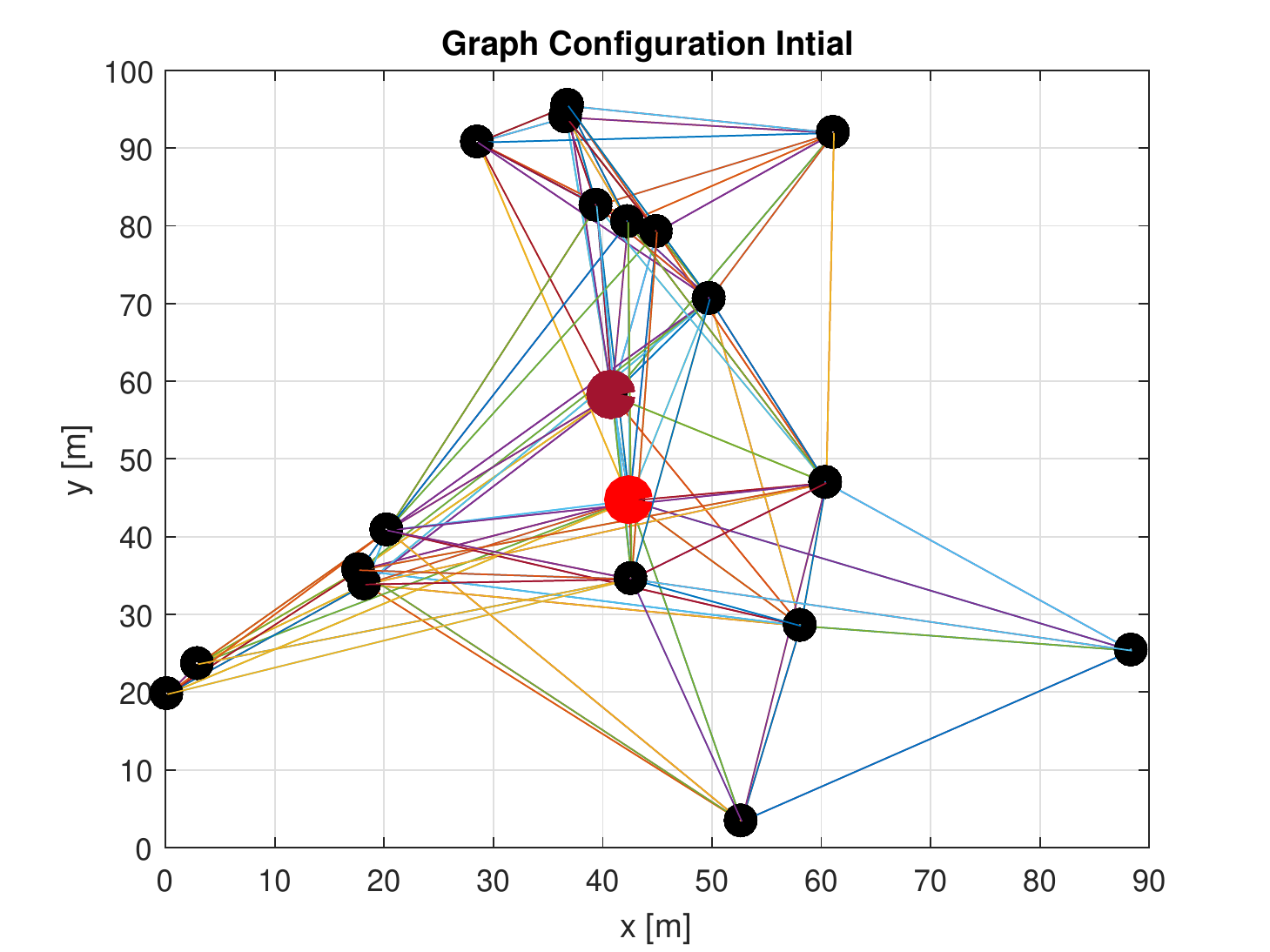}\label{WMSRCon2FIntial graph}}
\subfigure[Formation flight]{\includegraphics[width = .5\columnwidth]{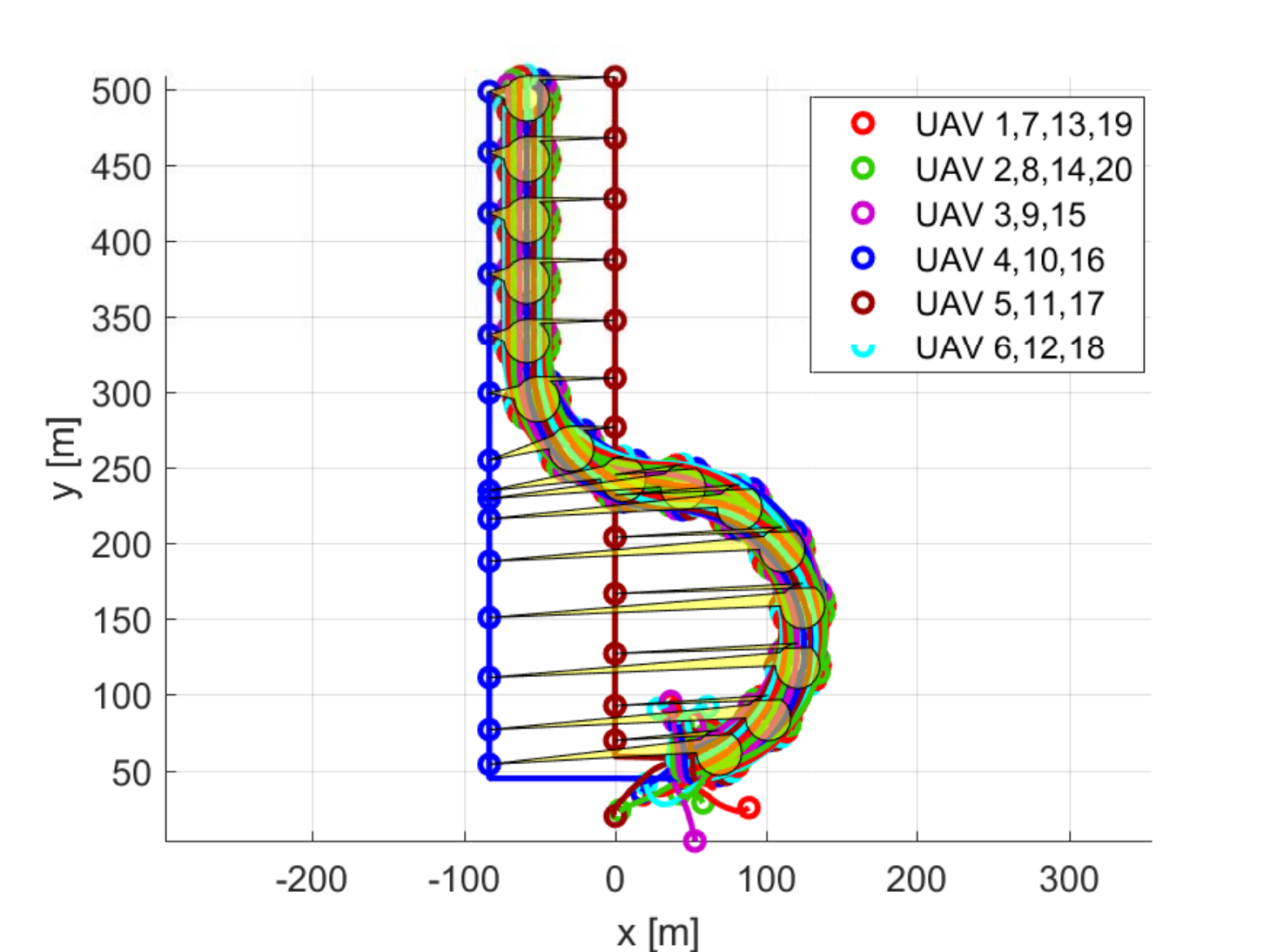}\label{WMSRCon2FFormation}}
\caption{(a) Initial random graph, with two misbehaving UAVs having  maximum communication link and maximum communication signal strength respectively ( red and dark red)    (b) Formation flight 20 UAVs in the presence of two (	$F=2$) misbehaving UAVs with a constant x position.}
\label{regularCons 1}
\end{figure}

\begin{figure}
\subfigure[x-velocity ]{\includegraphics[width = .5\columnwidth]{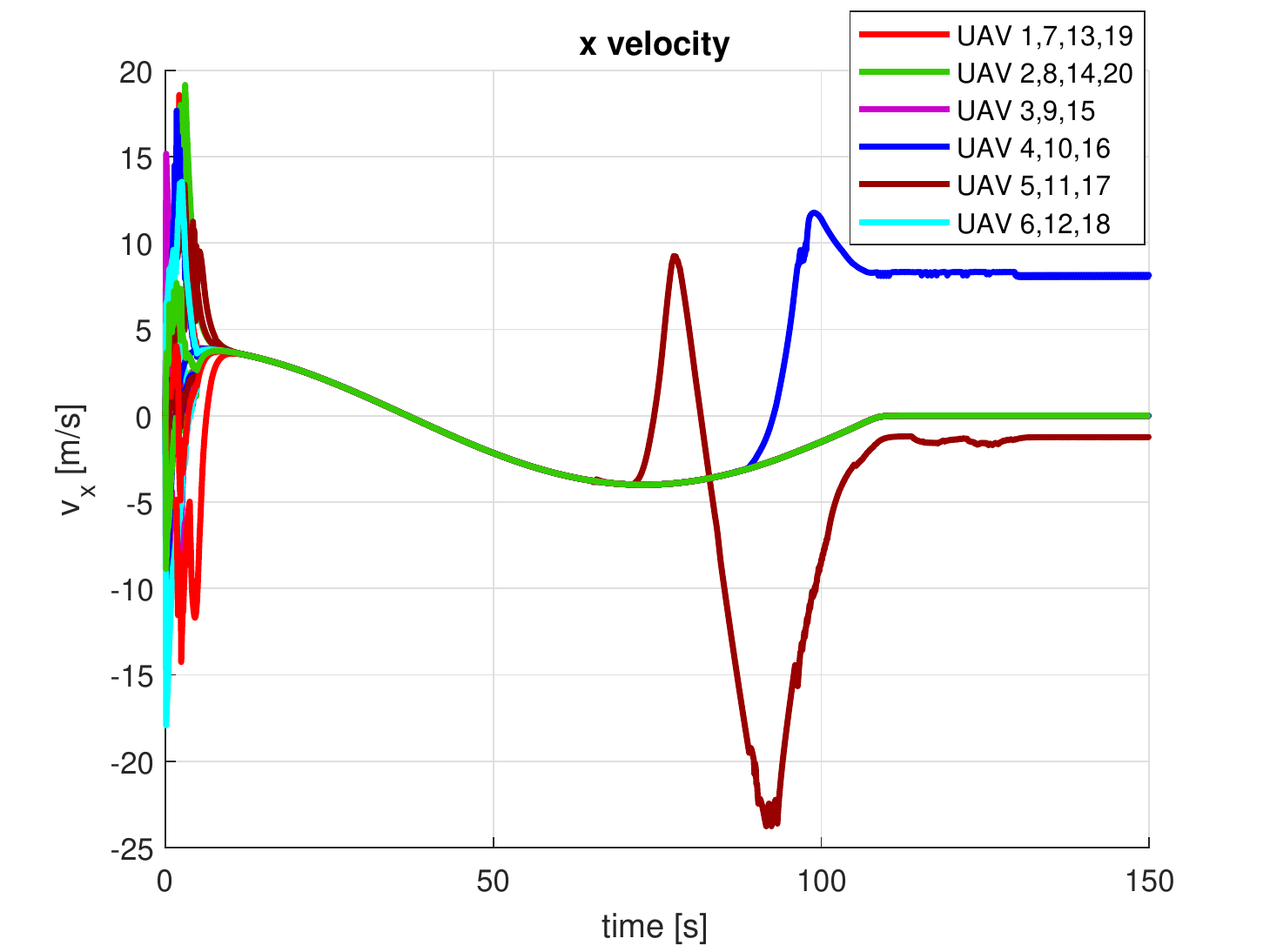}\label{WMSRCon2Fx-velo}}
\subfigure[x-position ]{\includegraphics[width = .5\columnwidth]{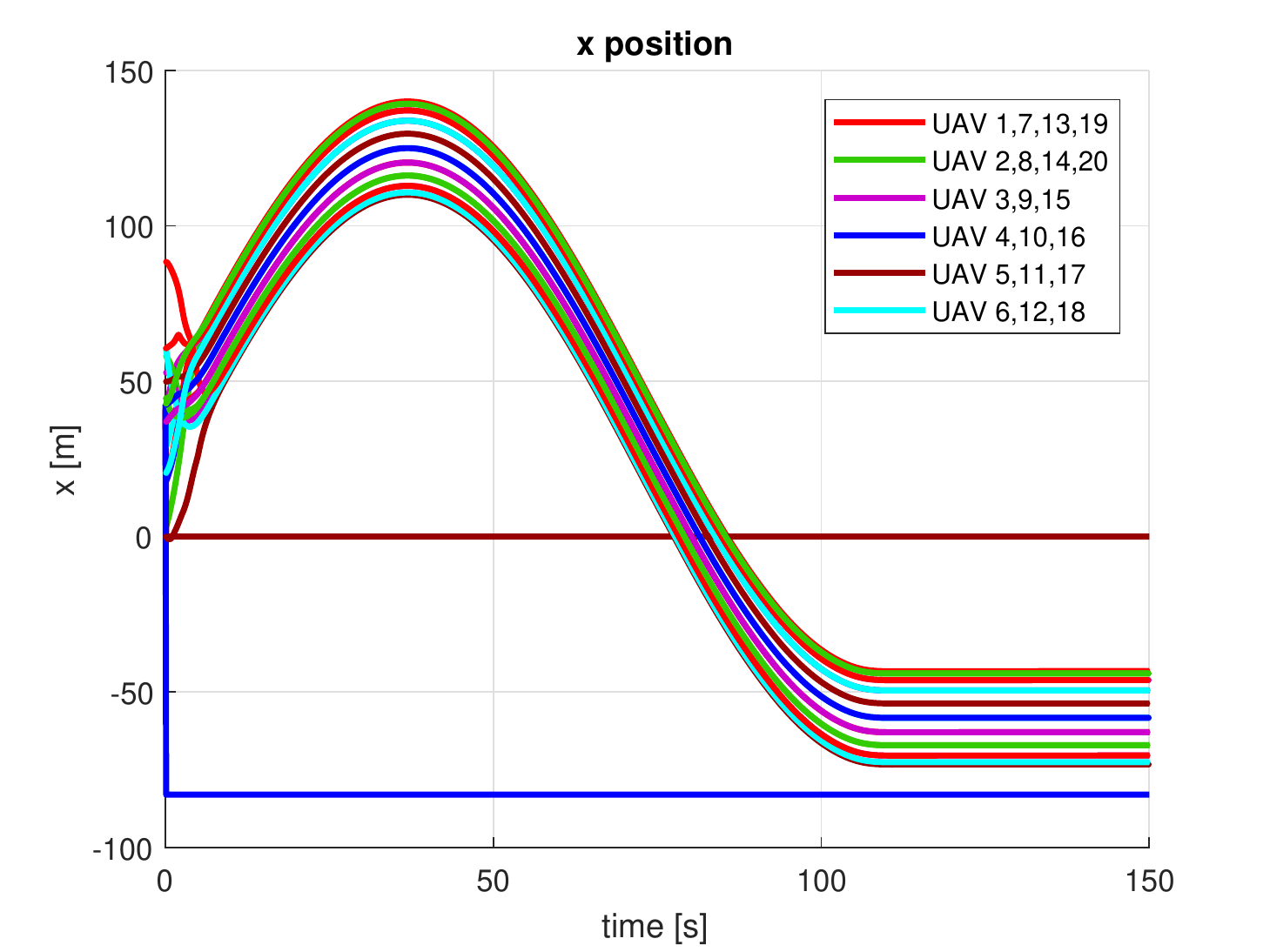}\label{WMSRCon2Fx-po}}
\caption{(a) x-direction velocities  (b) x-positions of the 20 UAVs under formation flight  where two of them are misbehaving UAVs.}
\label{regularCons 1}
\end{figure}

Here also a randomly generated graph in Figure \ref{WMSRCon2FIntial graph} with two (i.e., $F=2$) non-cooperative UAVs, a W-MSR algorithm based formation control applied with a connectivity controller. The two misbehaving UAVs keep a constant $x$ position with $x_{p_i}=-83m$ and $x_{p_j}=0, \forall i,j\in \mathcal{N}_a $, respectively as seen in the Figure \ref{WMSRCon2Fx-po}. These values are chosen purposely so that the misbehaving UAVs will be always in the communication range.

The cooperative UAVs converges to the formation flight within the initial value convex hull of the cooperative UAVs, despite the two non-cooperative UAVs constantly sharing a fixed position value to the W-MSR algorithm based formation control. To guarantee the W-MSR based formation control convergence, the algebraic connectivity of the network of UAVs should be kept above the resilience threshold, $\lambda_2 > 4F$,  i.e., $\lambda_2 > 8$.
\begin{figure}[h]
\subfigure[Final communication Graph ]{\includegraphics[width = .5\columnwidth]{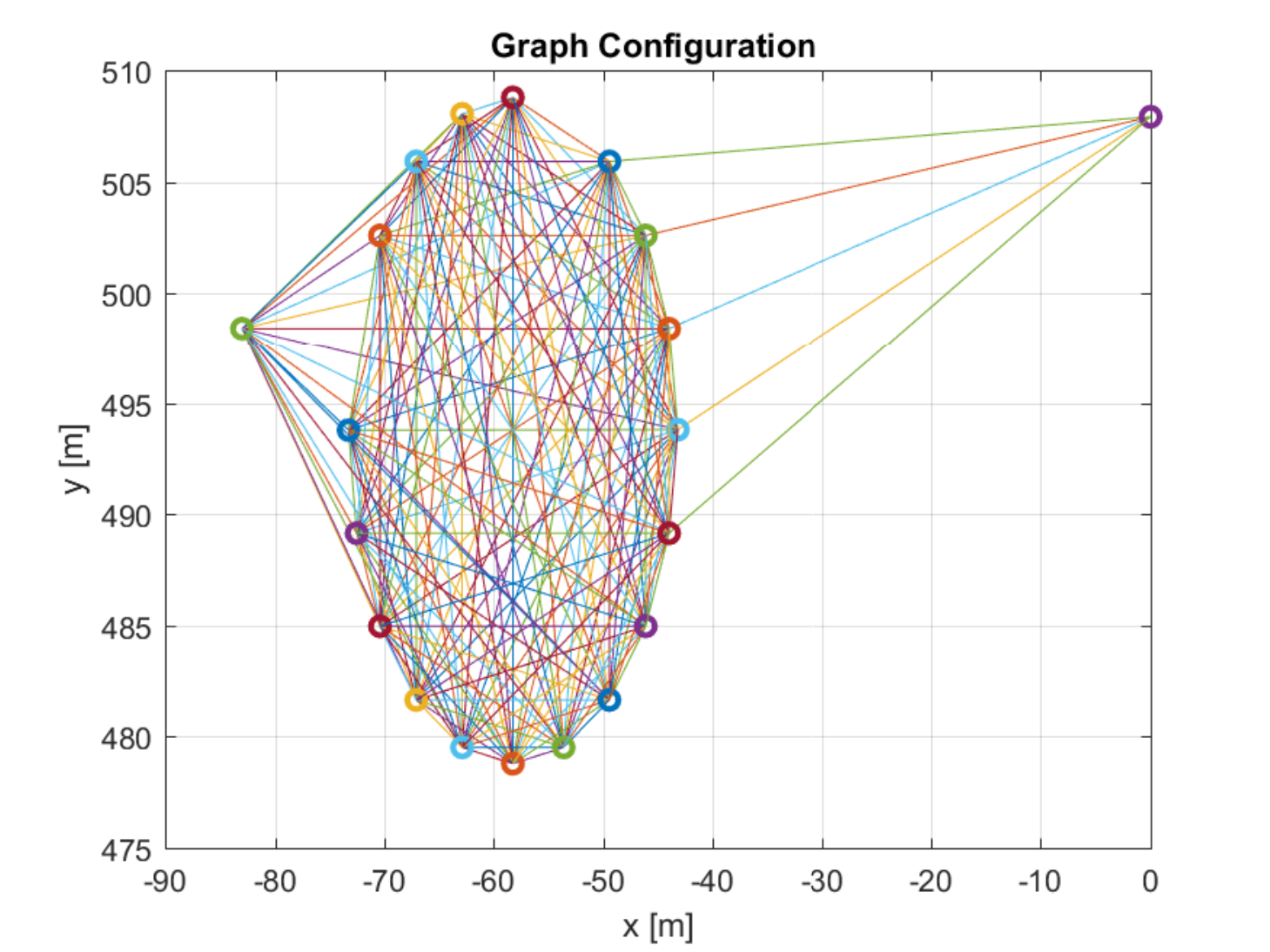}\label{WMSRCon2Ffinal graph}}
\subfigure[Algebric connectivity $\lambda_2(t)$  ]{\includegraphics[width = .5\columnwidth]{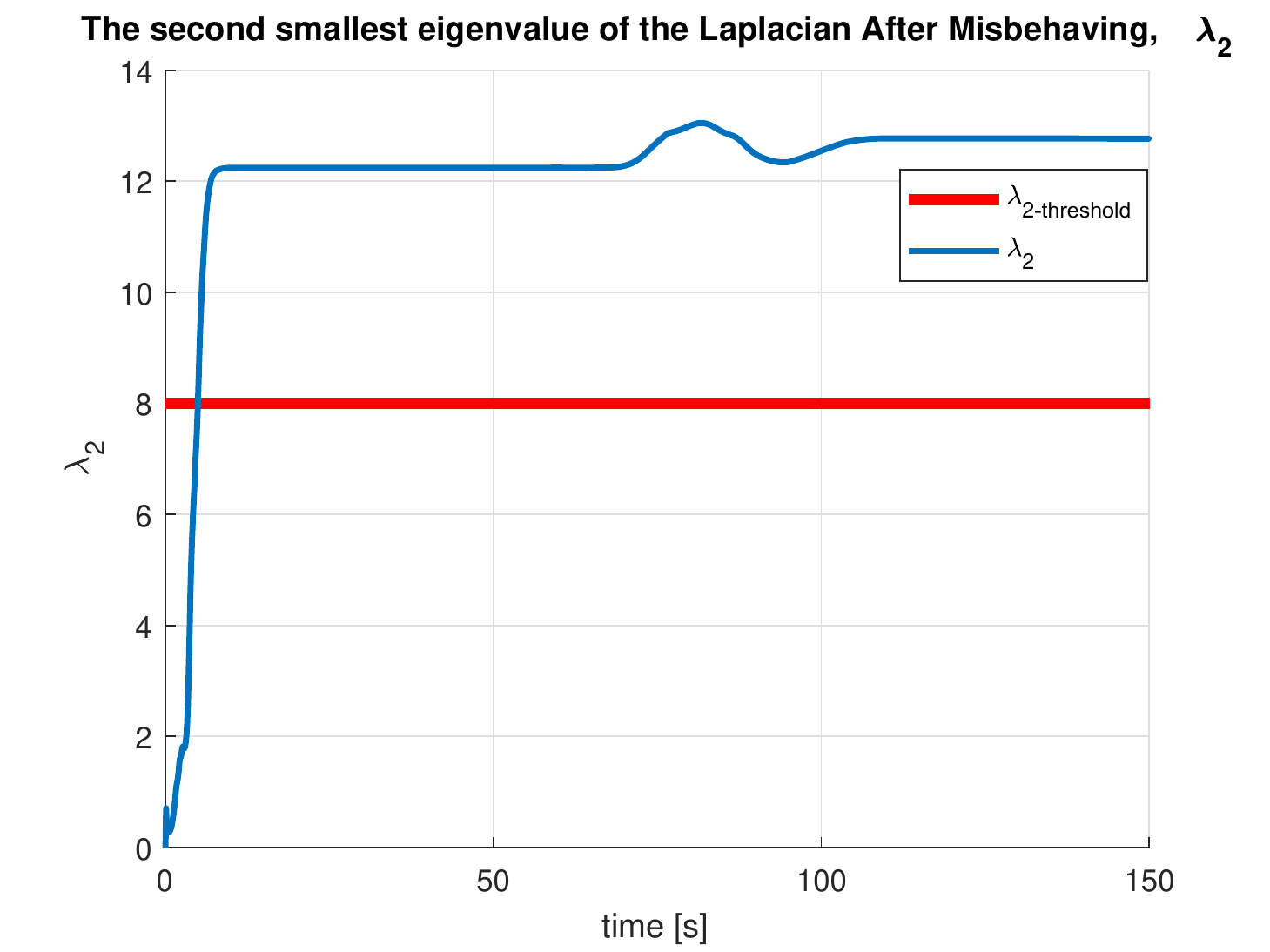}\label{WMSRCon2FseigneVal-2}}
\caption{(a)Final communication graph of the formation flying UAVs with W-MSR algorithm.  (b) Algebraic connectivity of the system of UAVs under formation flying kept above the threshold for resilience of W-MSR algorithm}.
\label{regularCons 1}
\end{figure}

\subsubsection{Misbehaving UAVs with offset in X-position}
\begin{figure}[H]
\subfigure[Intial Random Graph ]{\includegraphics[width = .5\columnwidth]{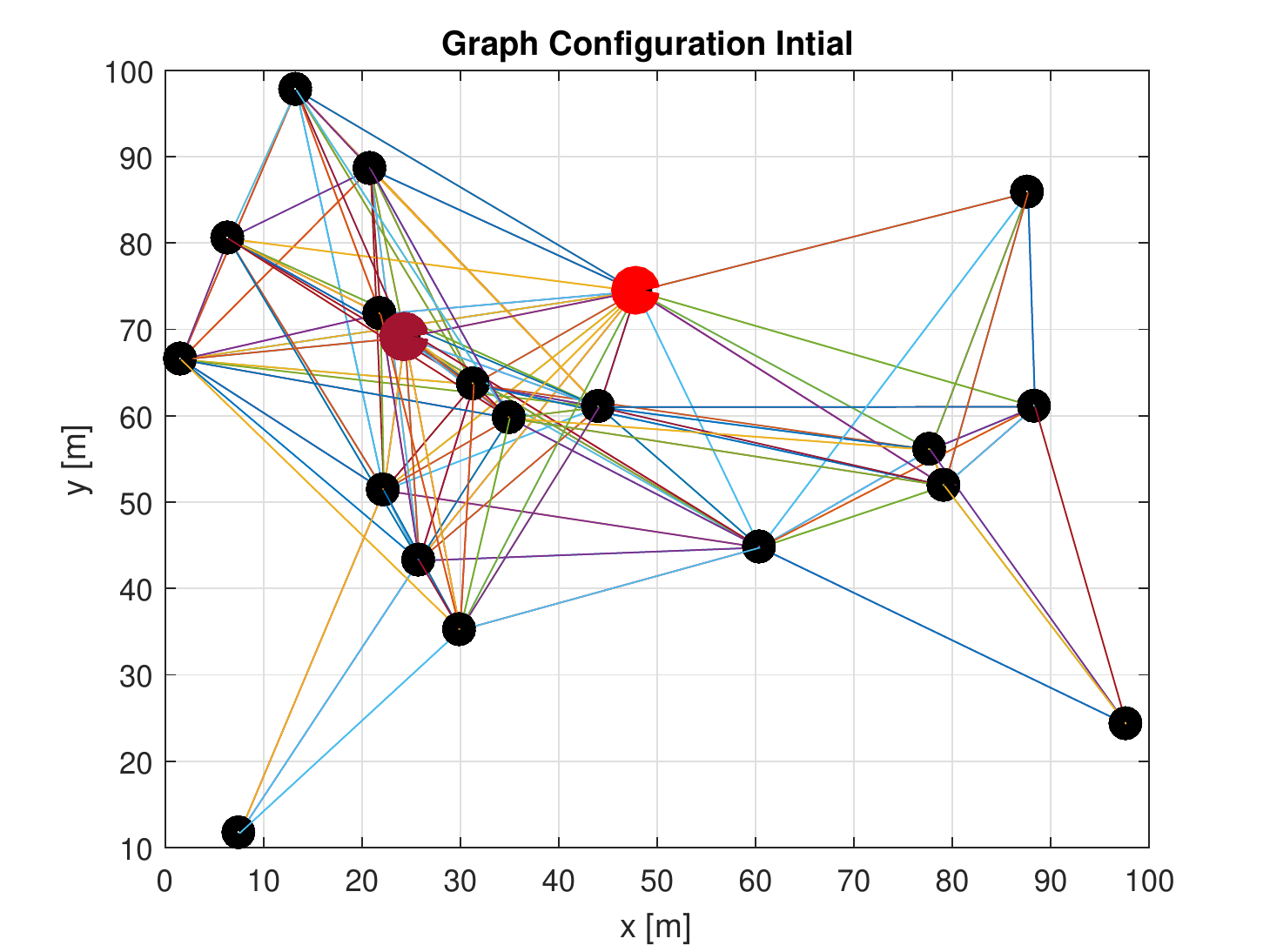}\label{WMSR2FOffSetIntial graph}}
\subfigure[Formation flight]{\includegraphics[width = .5\columnwidth]{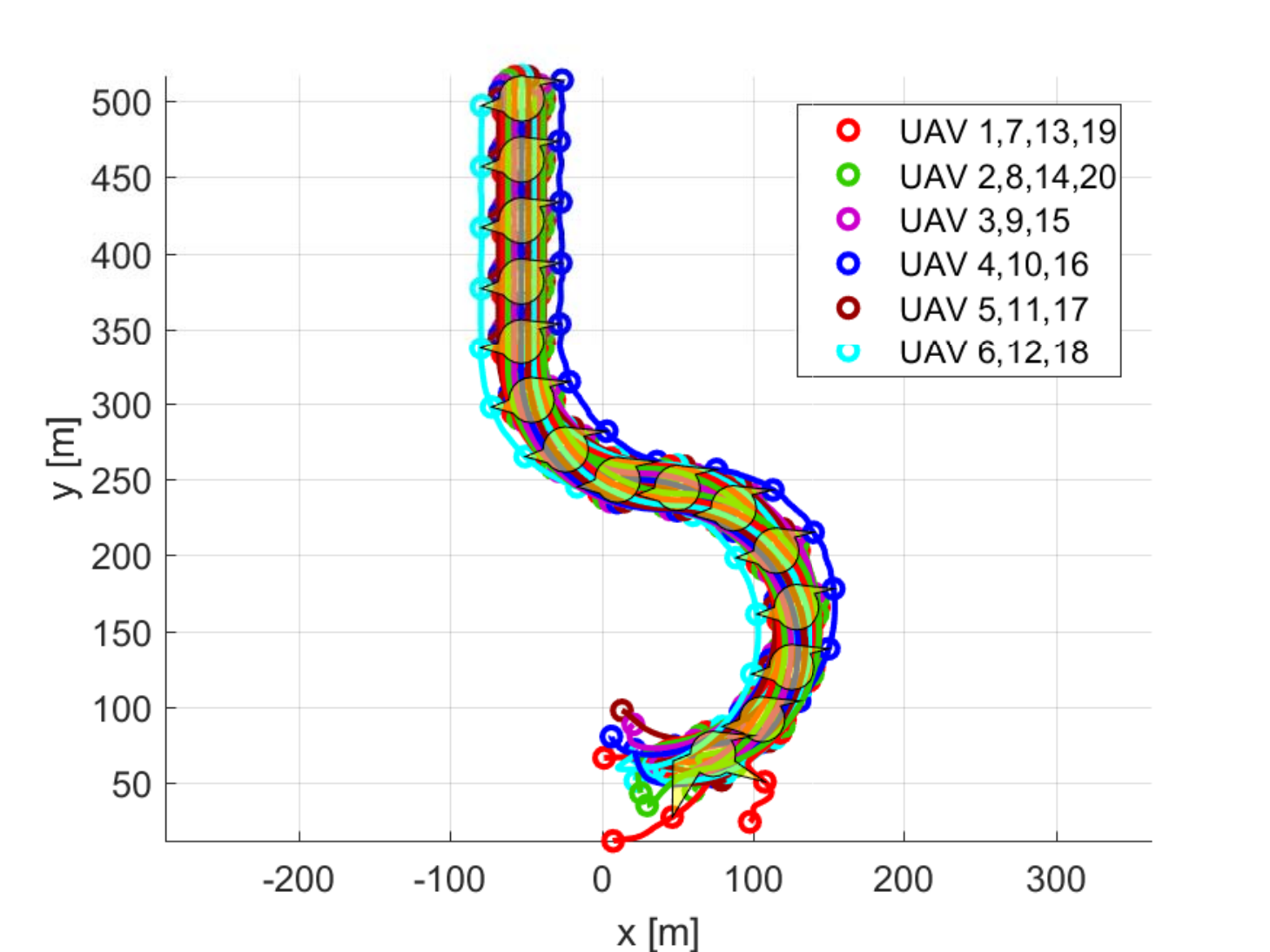}\label{WMSR2FOffSetFormation}}
\caption{(a) x-direction velocities  (b) y-direction velocities of the UAVs under formation fly.}
\label{regularCons 1}
\end{figure}

As indicated in the communication graph in the Figure \ref{WMSR2FOffSetIntial graph} the location of 20 UAVs are generated randomly within the communication range indicated in equation \ref{eq: communication function}. The two UAVs only maintain an offset in their x-position but kept the same heading and speed of the other UAVs, and consequently they stayed in the communication link range to influence the other well behaving UAVs.

\begin{figure}[H]
\subfigure[x-velocity ]{\includegraphics[width = .5\columnwidth]{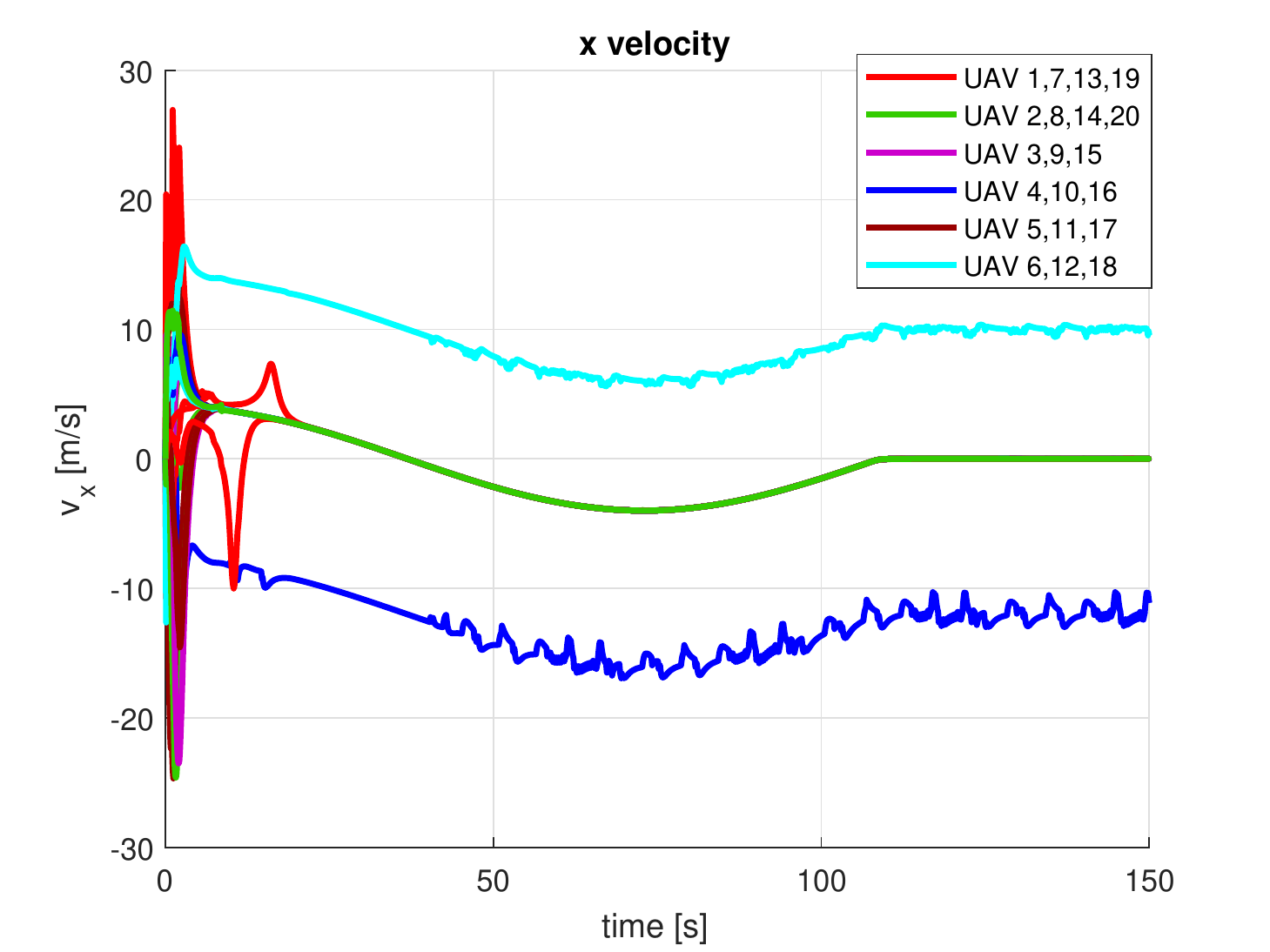}\label{WMSR2FOffSetx-velo}}
\subfigure[x-position ]{\includegraphics[width = .5\columnwidth]{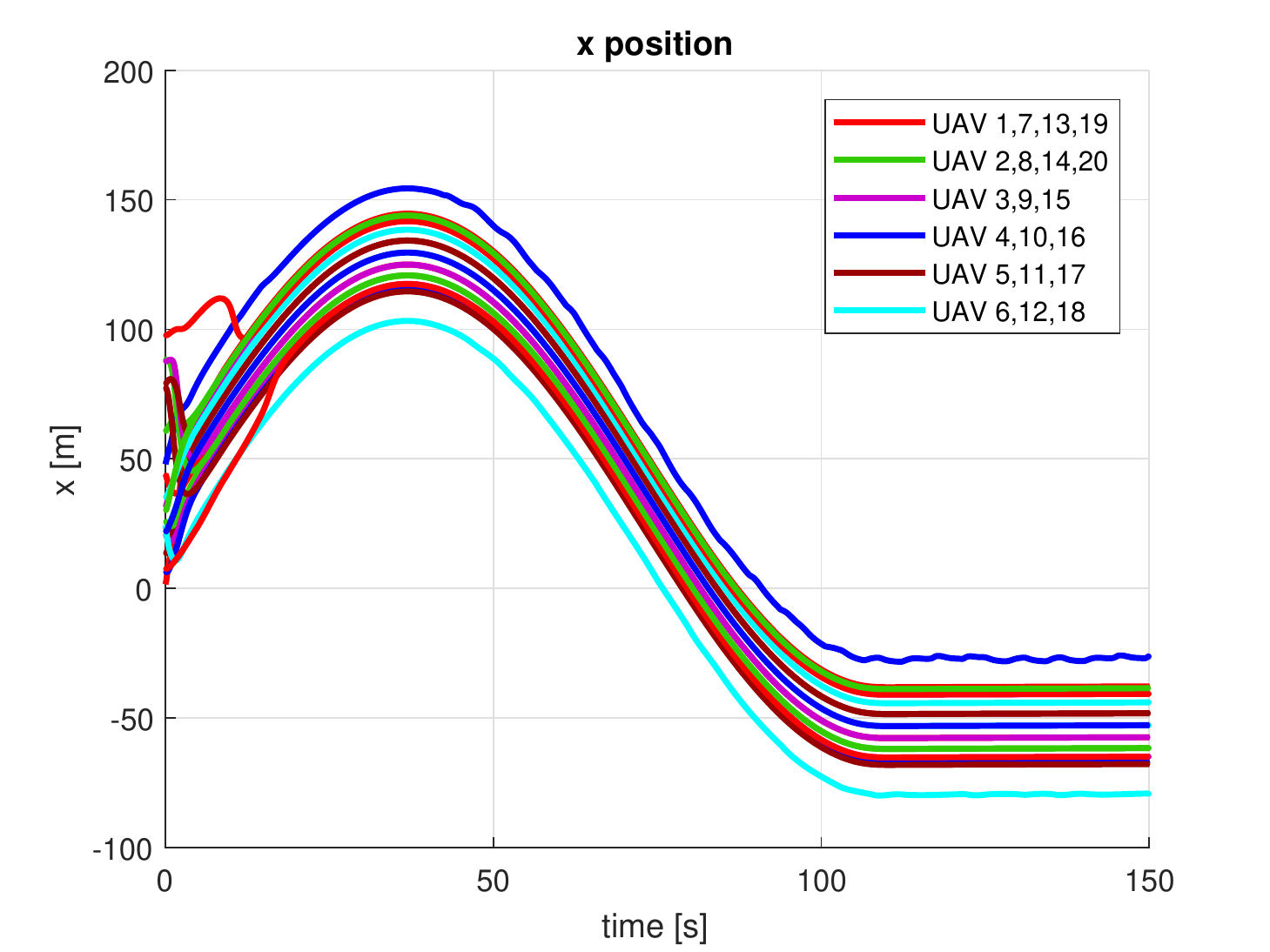}\label{WMSR2FOffSetx-po}}
\caption{(a) x-direction velocities  (b) x-direction velocities of the UAVs under formation fly.}
\label{regularCons 1}
\end{figure}

Throughout the flight, the cooperative UAVs converges within the initial value of the cooperative UAVs despite a constant influence of offset values sent by the two misbehaving UAVs. The cooperative UAVs maintain the formation and flight route (Figure \ref{WMSR2FOffSetFormation}) with the desired heading velocity (Figure \ref{WMSR2FOffSetx-velo} ) and heading direction(Figure \ref{WMSR2FOffSetx-po}).

\begin{figure}[H]
\subfigure[Final communication Graph ]{\includegraphics[width = .5\columnwidth]{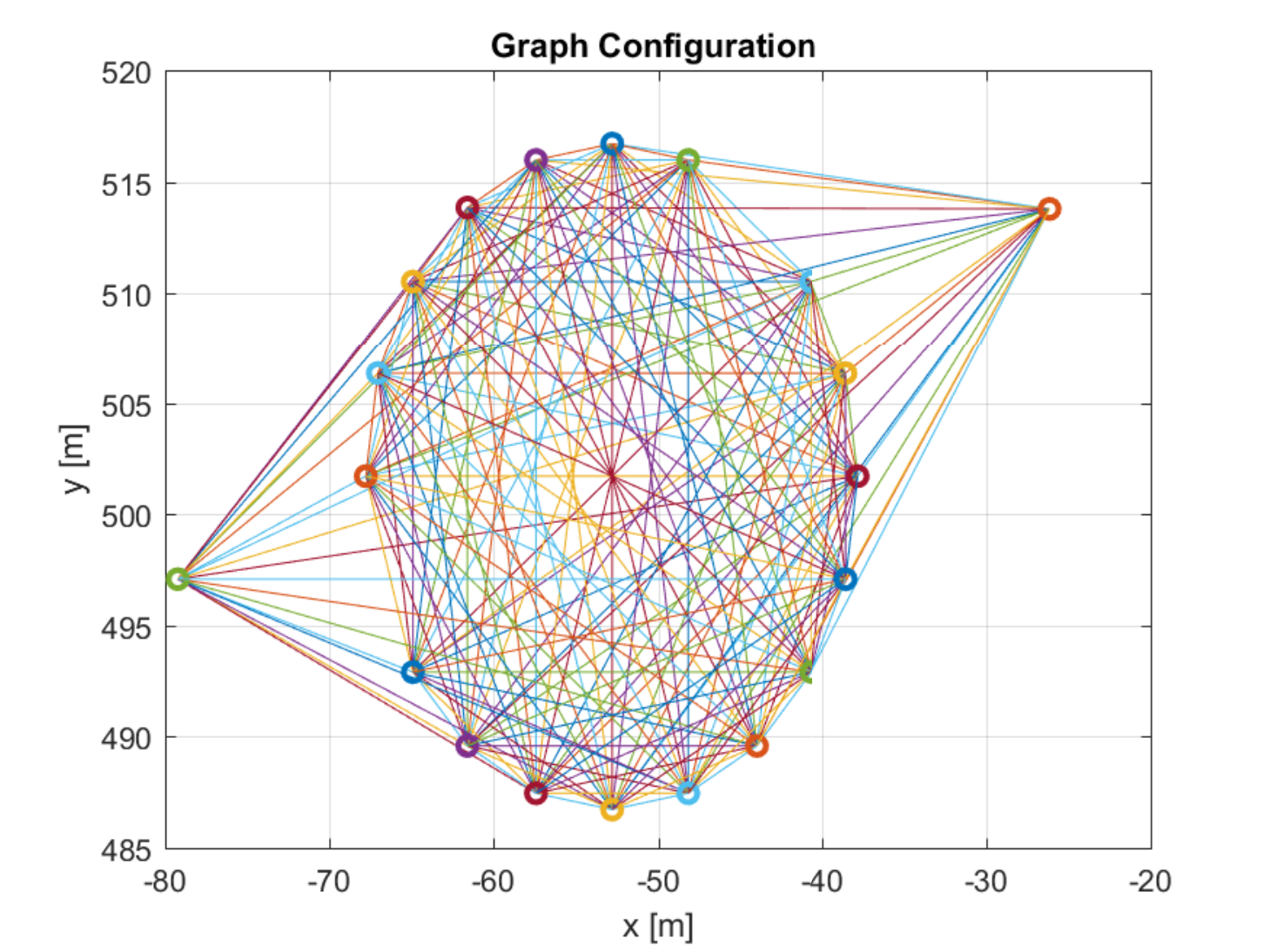}\label{WMSR2FOffSetfinal graph}}
\subfigure[Algebric connectivity $\lambda_2(t)$  ]{\includegraphics[width = .5\columnwidth]{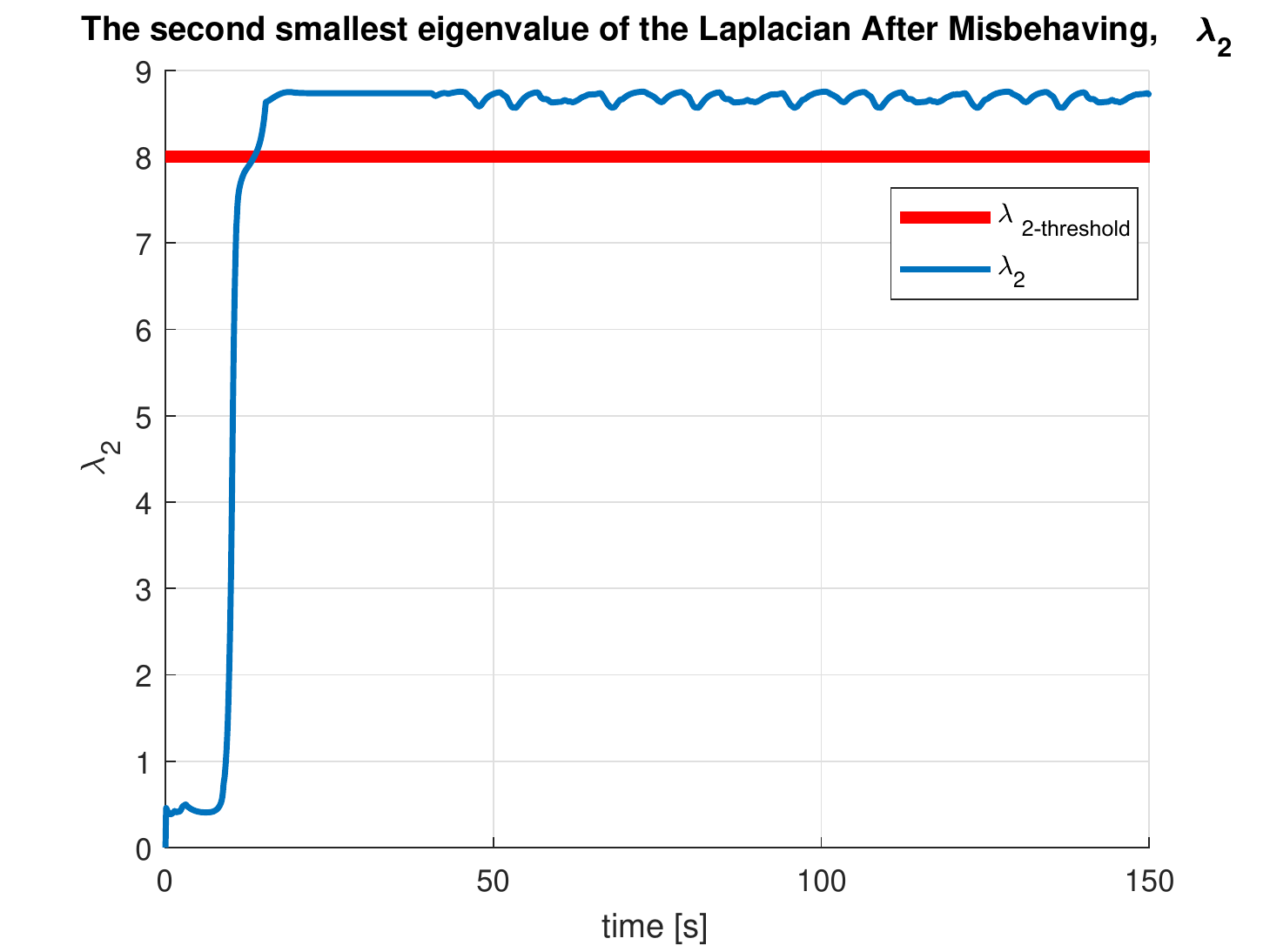}\label{WMSR2FOffSeteigneVal-2}}
\caption{(a) Final communication graph of the formation flying UAVs under W-MSR  consensus algorithm with 2 misbehaving UAVs.  (b) Algebraic connectivity of the system of UAVs system of UAVs.}
\label{regularCons 1}
\end{figure}

\subsubsection{ Misbehaving UAVs with Time Varying Signal}
\begin{figure}[H]
\subfigure[Intial Random Graph ]{\includegraphics[width = .5\columnwidth]{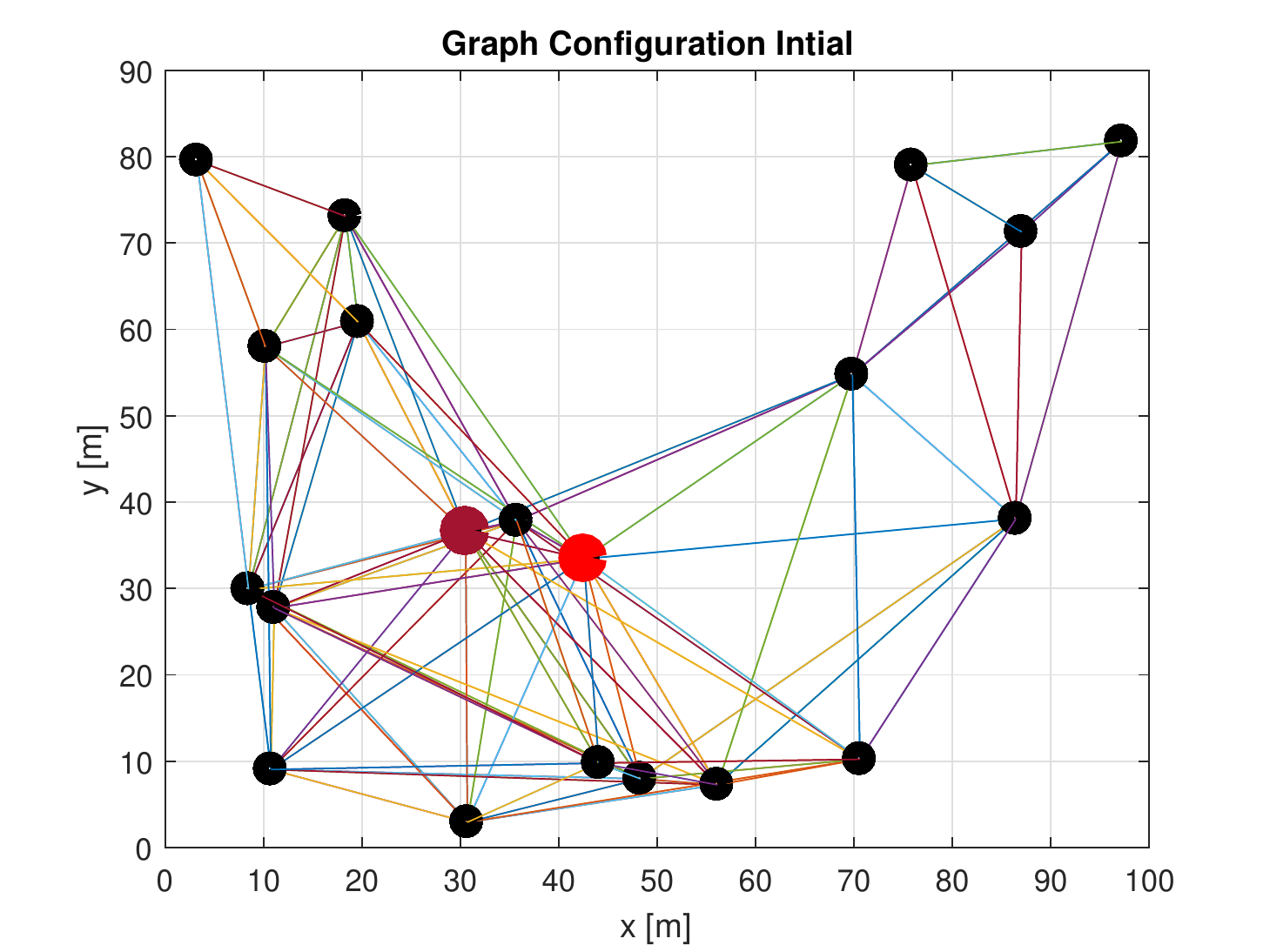}\label{WMSR2FTimeVarIntial graph}}
\subfigure[Formation flight]{\includegraphics[width = .5\columnwidth]{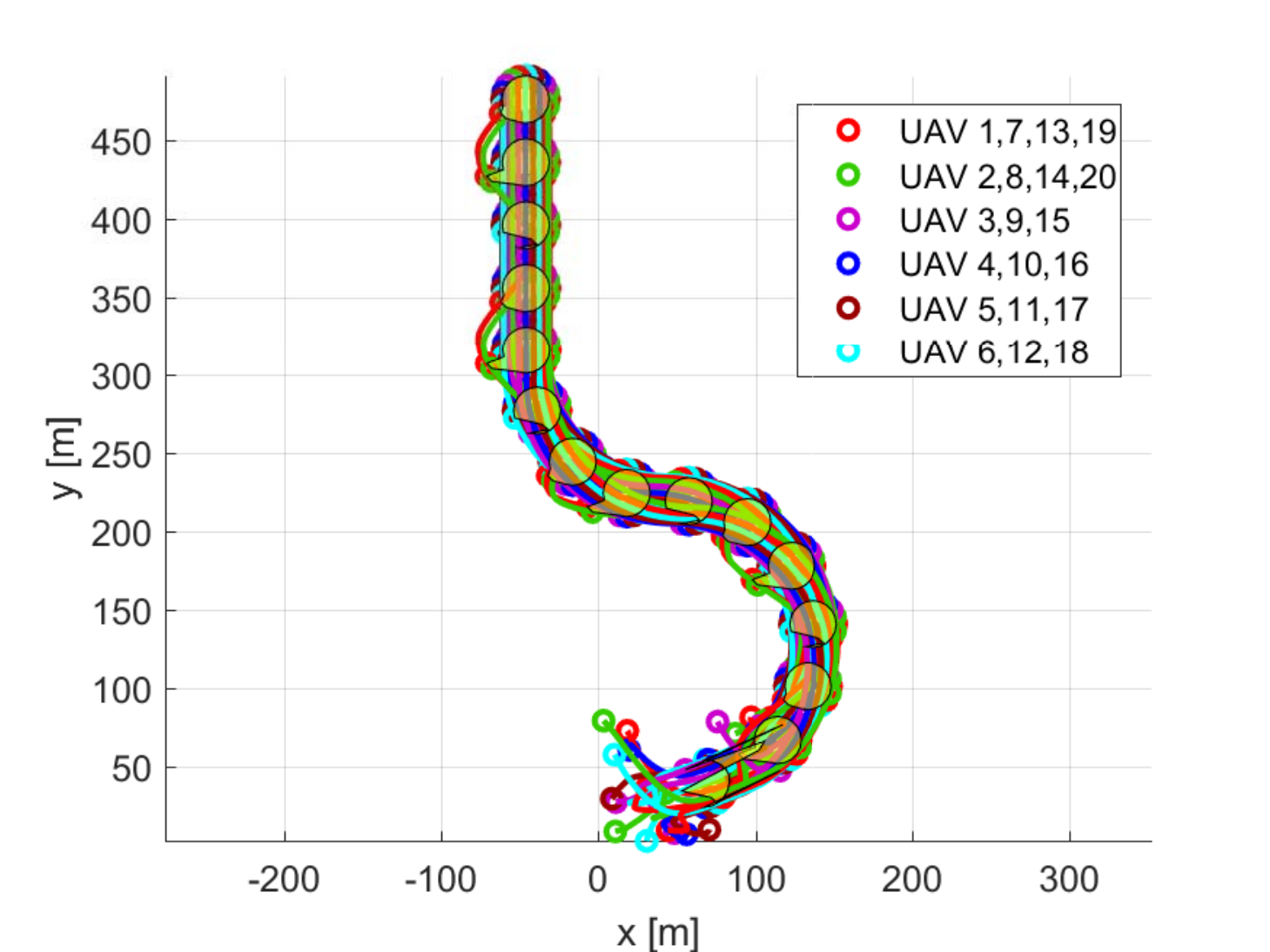}\label{WMSR2FTimeVarFormation}}
\caption{(a) Initial random graph  (b) Formation flight the UAVs in the presence of two UAVs sharing time varying heading.}
\label{WMSR2FTimeVar Formation }
\end{figure}

In this scenario, the noncooperating UAVs shares a time-varying heading and kept close to the other UAVs to maintain a communication link.  For a randomly generated graph in Figure \ref{WMSR2FTimeVarIntial graph} with two  noncooperating UAVs, a W-MSR algorithm based formation flight applied with a connectivity controller. It is clear that despite the two non cooperating UAVs who maintains a time-varying heading, the other well-behaving UAVs maintained the formation flight with the correct route (Figure \ref{WMSR2FTimeVarFormation}).

\begin{figure}[H]
\subfigure[x-velocity ]{\includegraphics[width = .5\columnwidth]{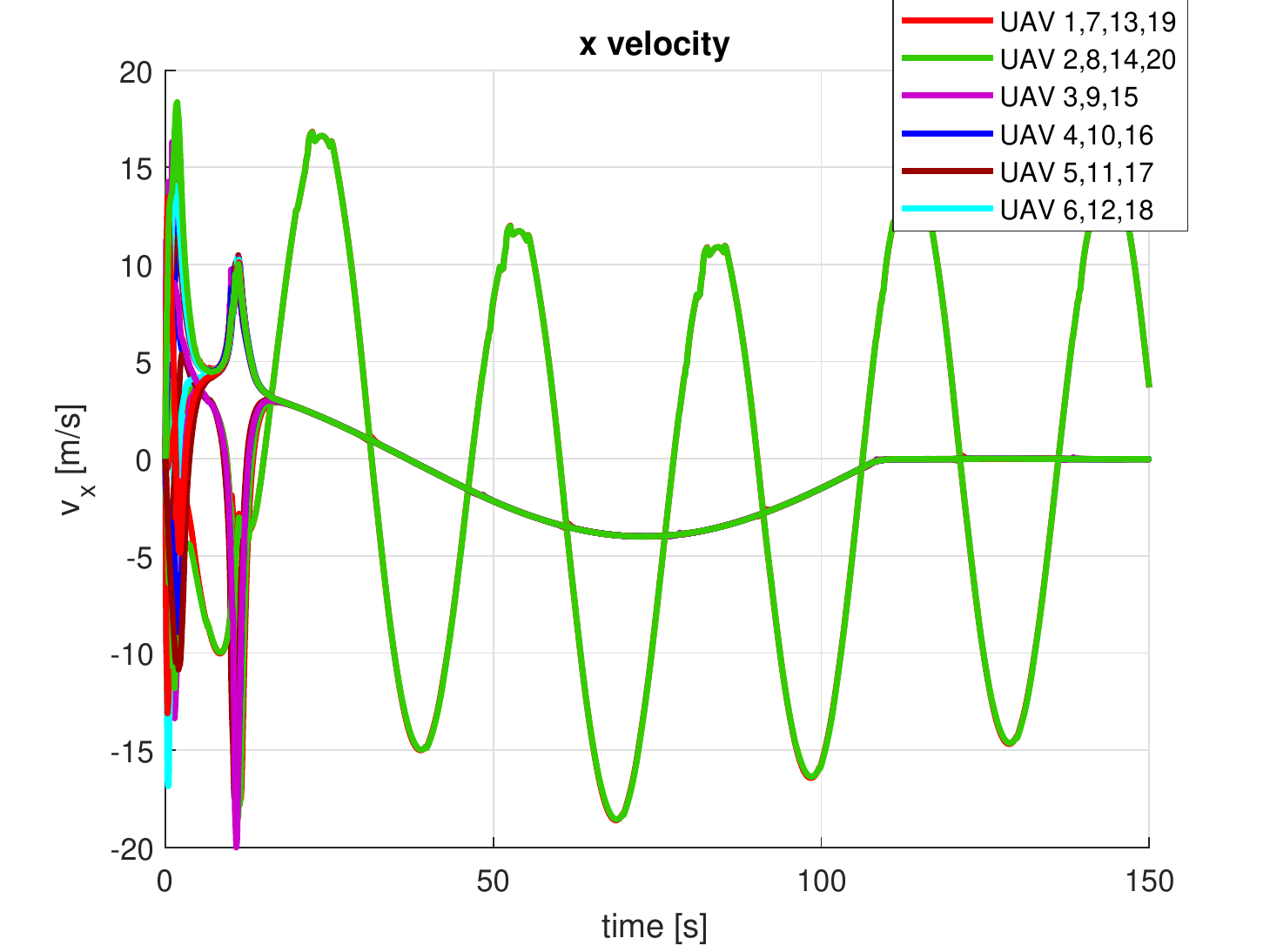}\label{WMSR2FTimeVarx-velo}}
\subfigure[x-position ]{\includegraphics[width = .5\columnwidth]{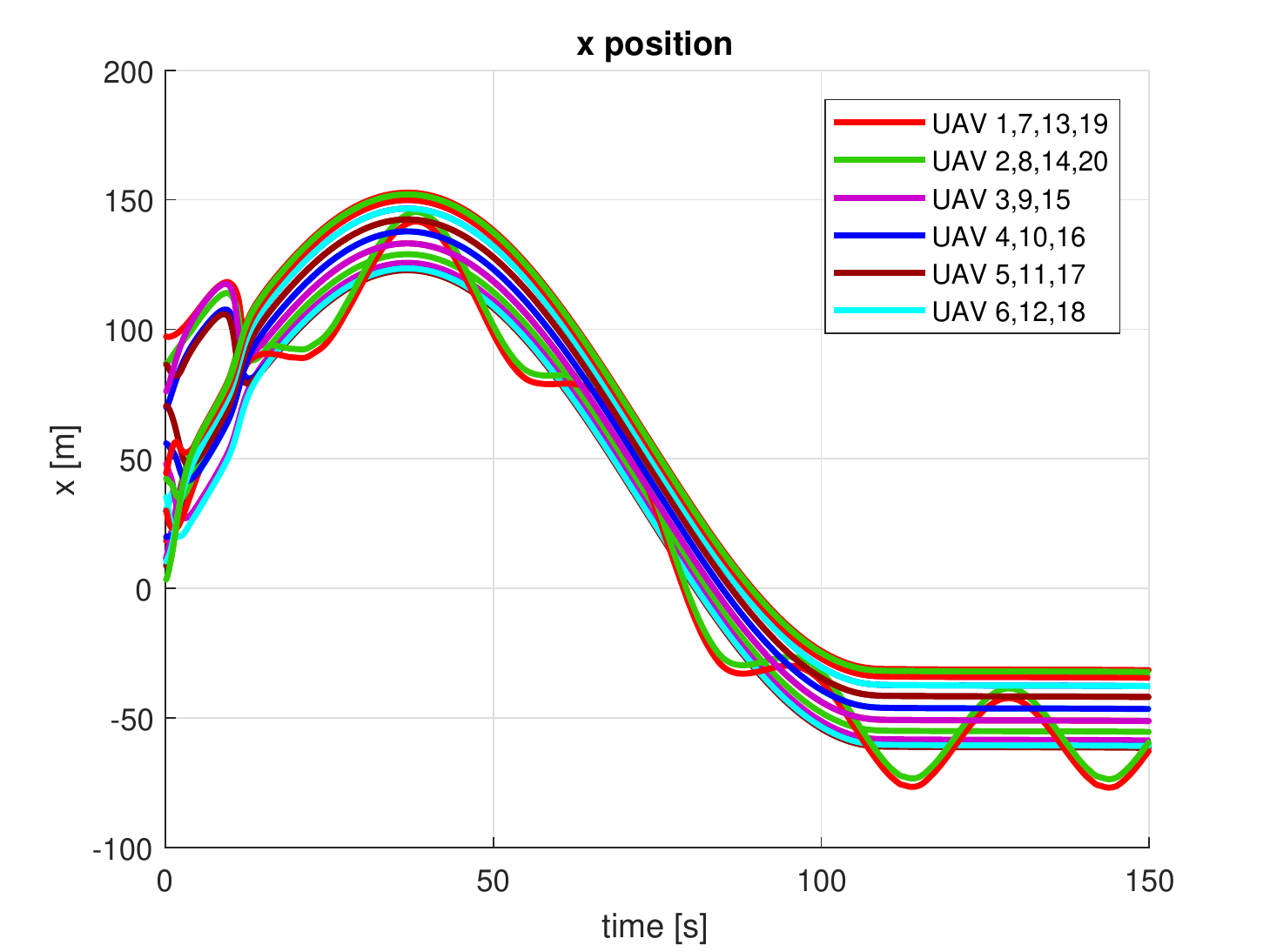}\label{WMSR2FTimeVarx-po}}
\caption{(a) x-direction velocities  (b) x-positions of the UAVs under formation fly in the presence of two miss behaving UAVs with time varying heading.}
\label{WMSR2FTimeVar X}
\end{figure}
The trajectories indicate that the well behaving UAVs achieved consensus both in their velocity and position for the formation flight. They maintain the desired constant heading velocity ($[v_x=0, v_y=4m/s]^T $)  (Figure \ref{WMSR2FTimeVar X}) at the last part of the designated route while flying in a formation.

\begin{figure}
\subfigure[Final communication Graph ]{\includegraphics[width = .5\columnwidth]{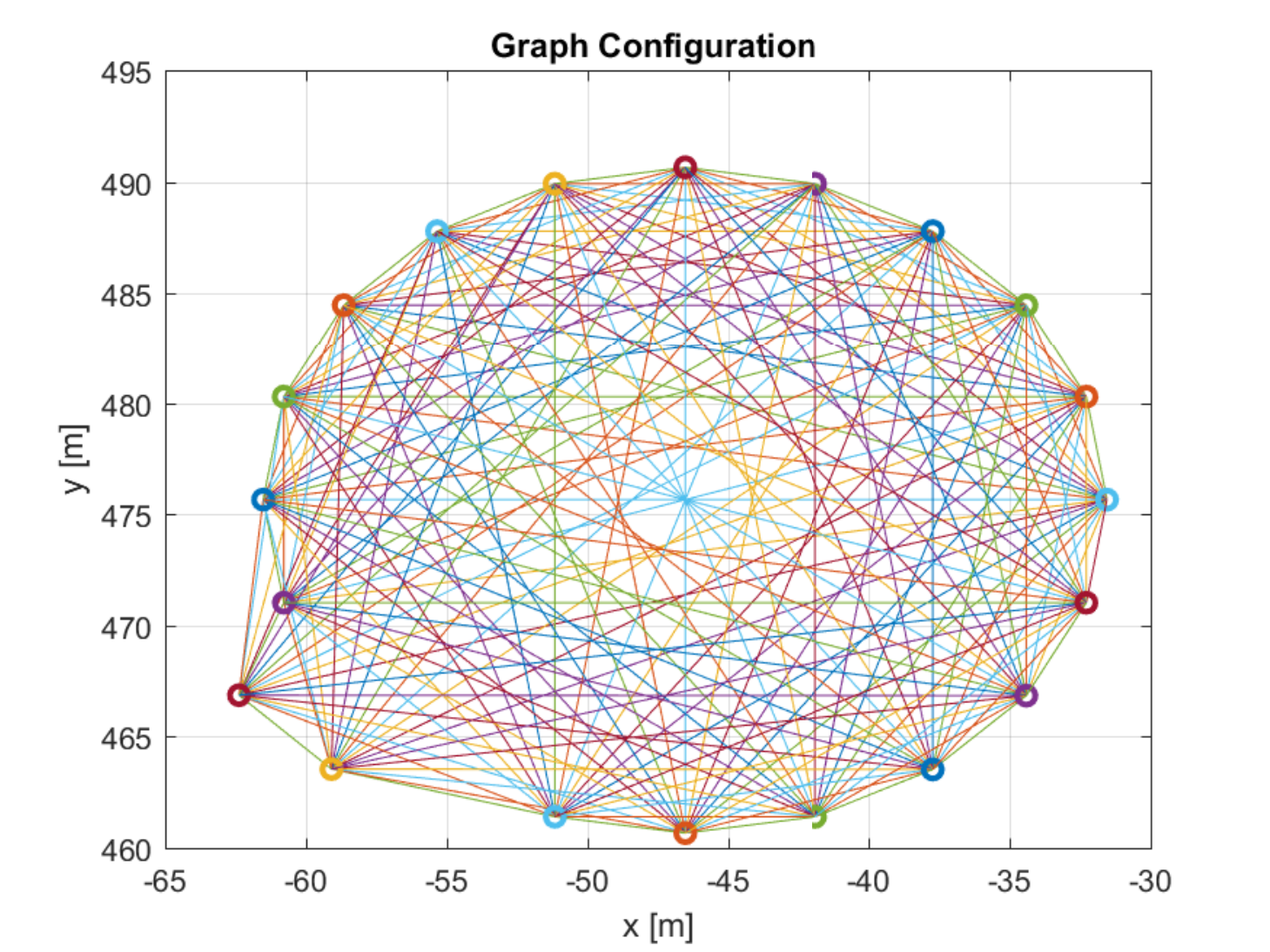}\label{WMSR2FTimeVarfinal graph}}
\subfigure[Algebraic connectivity $\lambda_2(t)$  ]{\includegraphics[width = .5\columnwidth]{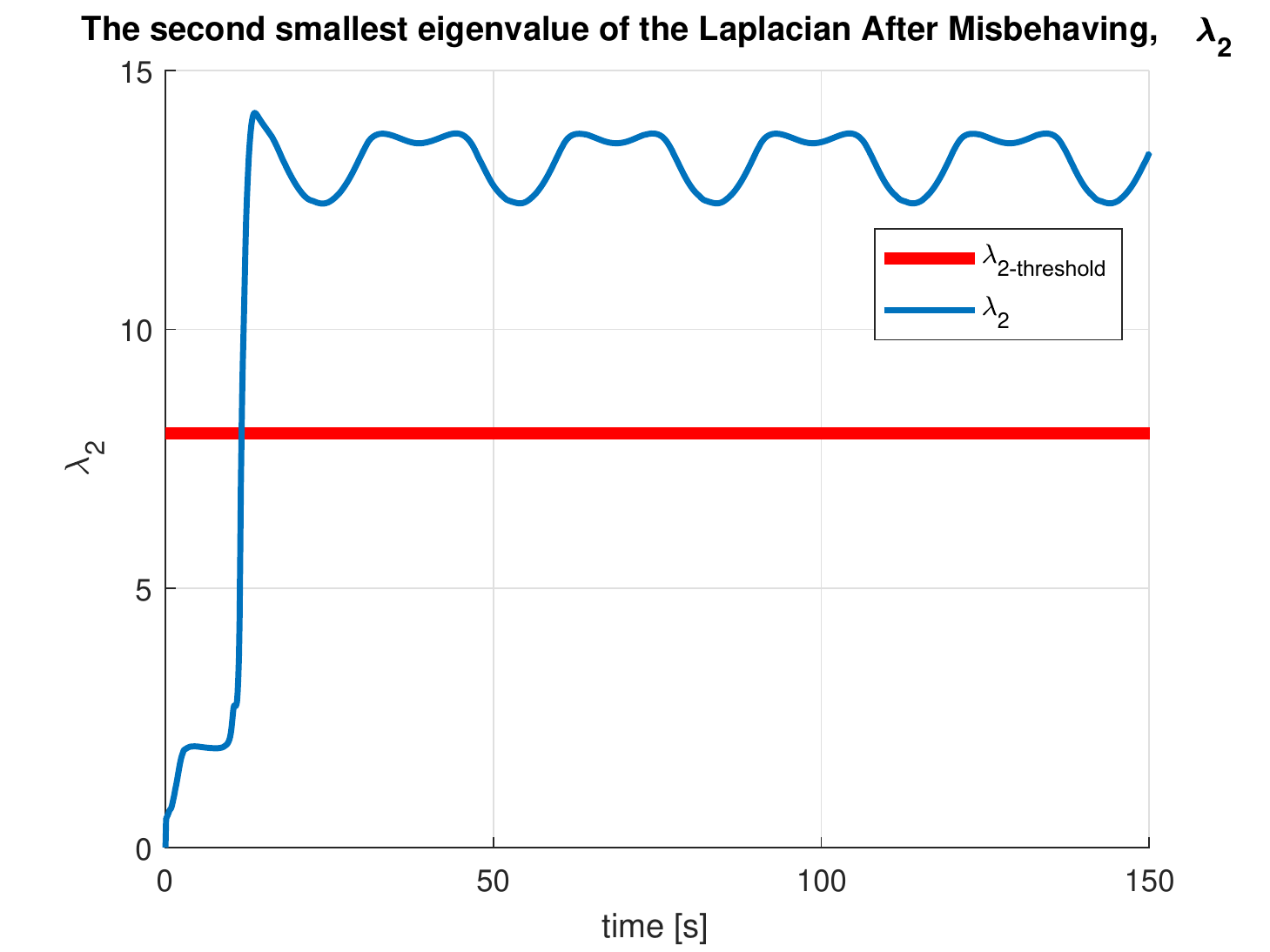}\label{WMSR2FTimeVareigneVal-2}}
\caption{(a) Final communication graph of the formation flying UAVs under W-MSR  consensus algorithm with 2 misbehaving UAVs.  (b) Algebraic connectivity of the system of UAVs system of UAVs.}
\label{WMSR2FTimeVar Eig}
\end{figure}

The final communication graph in Figure \ref{WMSR2FTimeVarfinal graph} shows that all the 20 UAVs in the formation flight stayed connected despite the the two UAVs are constantly sharing a varying position offset with their neighbors as
\begin{equation} \label{time vaying offset}
x_{p_i}(t+1)=x_{p_i}(t)+sin(t), \quad  \forall i\in \mathcal{N}_a
\end{equation}

and the UAVs still achieve resilience. But to guarantee the W-MSR based formation flight consensu convergence, the controller keeps the algebraic connectivity of the communication graph above the threshold as seen in figure \ref{WMSR2FTimeVareigneVal-2}

\section{conclusion}
In this work, we studied resilient consensus problem for a network of UAVs with second-order dynamics where the maximum number of misbehaving UAVs in the network is known. A W-MSR algorithm is proposed for well-behaving UAVs to fly in a formation keeping the desired route. The resilient consensus is guaranteed on the  $(2F+1)$-robustness of the network topology where $r$-robustness is a local information. A gradient based distributed controller is designed with a power iteration method to estimate the second eigenvector.

The misbehaving UAVs assume from constant to time-varying heading and also an offset on their true positions to depict a fault in the sensors. In all the models of the misbehaving UAVs, the formation control with W-MSR algorithm managed to maintain resilient consensus. The controllers made sure they maintain the algebraic connectivity of the underlying graph above the resilient threshold to guarantee convergence. We also noticed that the normal consensus algorithm fails to be resilient in the presence of misbehaving UAVs while even the algebraic connectivity was well kept in check showing the traditional metric like graph connectivity is not a measure of  resiliency of a network. Whereas algorithms with local information, i.e., r-robustness guarantee well behaving UAVs to achieve consensus in the presence of misbehaving UAVS.

	\section{Acknowledgments}
	
	This work was supported in part by ICT R\&D program of MSIP/IITP [R-20150223-000167, Development of High Reliable Communications and Security SW for Various Unmanned Vehicles].

\iftrue

\fi

\end{document}